\documentclass[aps,preprint,eqsecnum,showpacs,nofootinbib]{revtex4}
\usepackage{epsfig}
\begin{document}

\title{Next-to-leading order QCD predictions for pair production
of neutral Higgs bosons at the CERN Large Hadron Collider}

\author{Li Gang Jin}
 \affiliation{Institute of Theoretical Physics, Academia
Sinica, Beijing 100080, China}

\author{Chong Sheng Li\footnote{\hspace{-0.1cm}Electronics address:
csli@pku.edu.cn}, Qiang Li, and Jian Jun Liu} \affiliation{
Department of Physics, Peking University, Beijing 100871, China}

\author{Robert J. Oakes}
\affiliation{Department of Physics and Astronomy, Northwestern
University, Evanston, IL 60208-3112, USA}
\date{\today}

\begin{abstract}
We present the calculations of the complete NLO inclusive total
cross sections for pair production of neutral Higgs bosons through
$b\bar b$ annihilation in the minimal supersymmetric standard
model at the CERN Large Hadron Collider. In our calculations, we
used both the DREG scheme and the DRED scheme and found that the
NLO total cross sections in these two schemes are the same. Our
results show that the $b\bar b$-annihilation contributions can
exceed those of $gg$ fusion and $q\bar q$ annihilation for
$h^0H^0$, $A^0h^0$ and $A^0H^0$ productions when $\tan\beta$ is
large. In the case of $\mu>0$, the NLO corrections enhance the LO
total cross sections significantly, reaching a few tens of
percent, while for $\mu<0$, the corrections are relatively small,
and are negative in most of parameter space. Moreover, the NLO QCD
corrections reduce the dependence of the total cross sections on
the renormalization/factorization scale, especially for $\mu<0$.
We also used the CTEQ6.1 PDF sets to estimate the uncertainty of
LO and NLO total cross sections, and found that the uncertainty
arising from the choice of PDFs increases with the increasing
$m_{A^0}$.
\end{abstract}

\maketitle

\section{Introduction}
The Higgs mechanism plays a key role for spontaneous breaking of
the electroweak symmetry both in the standard model (SM) and in
the minimal supersymmetric (MSSM) extension of the SM
\cite{nilles}. Therefore, the search for Higgs bosons becomes one
of the prime tasks in future high-energy experiments, especially
at the CERN Large Hadron Collider (LHC), with $\sqrt{S}=14$ TeV
and a luminosity of 100 ${\rm fb^{-1}}$ per year \cite{lhc}. In
the SM, only one Higgs doublet is introduced, and the neutral
CP-even Higgs boson mass is basically a free parameter with a
theoretical upper bound of $m_H\leq600$ -- 800 GeV \cite{massh}
and a LEP2 experimental lower bound of $m_H\geq 114.4$ GeV
\cite{parameter}. In the MSSM, two Higgs doublets are required in
order to preserve supersymmetry (SUSY), and consequently the model
predicts five physical Higgs bosons: the neutral CP-even ones
$h^0$ and $H^0$, the neutral CP-odd one $A^0$, and the charged
ones $H^\pm$. The $h^0$, which behaves like the SM one in the
decoupling region ($m_{A^0}\gg m_{Z^0}$), is the lightest, and its
mass is constrained by a theoretical upper bound of $m_{h^0}\leq
140$ GeV when including the radiative corrections \cite{massh0}.
The analyses in \cite{detect} indicate that the $h^0$ boson can
not escape detection at the LHC, and that in large areas of the
parameter space, more than one Higgs particle in the MSSM can
possibly be found, which is an exciting result, since the
discovery of any additional Higgs bosons will be direct evidence
of physics beyond the SM.

At the LHC, a neutral Higgs boson $\phi$ can be produced through
following mechanisms: gluon fusion $gg \rightarrow \phi$
\cite{gg2h}, weak boson fusion $qq \rightarrow qqV^*V^*
\rightarrow qqh^0/qqH^0$ \cite{vv2h}, associated production with
weak bosons \cite{v2vh}, pair production \cite{hpair0,hpair1,
hpair11,hpair2}, and associated production with a $t\bar t$ pair
$gg/q\bar{q} \rightarrow t\bar{t}\phi$ \cite{ttbbh}. In the MSSM,
since the couplings between Higgs bosons and $b$ quarks can be
enhanced by large values of $\tan\beta$, the ratio of the vacuum
expectation values of the two Higgs doublets, Higgs bosons will
also be copiously produced in association with $b$ quarks at the
LHC. Except for $q\bar q \rightarrow b\bar b h^0$, the other
relevant production mechanisms depend on the final state being
observed \cite{differenth}. For inclusive Higgs production, the
lowest order process is $b\bar b\rightarrow h^0$ \cite{bbh}, and
the convergence of the perturbative expansion is improved by
summing the collinear logarithms to all orders through the use of
$b$ quark parton distributions with an appropriate factorization
scale. However, if at least one high-$p_T$ $b$ quark is required
to be observed, the leading partonic process is $gb\rightarrow
bh^0$ \cite{gbbh}, and if two high-$p_T$ $b$ quarks are required,
the leading subprocess is $gg\rightarrow b\bar b h^0$
\cite{ggbbh}.

Studying the pair production of neutral Higgs bosons may be an
important way to probe the trilinear neutral Higgs boson
couplings, which can distinguish between the SM and the MSSM. In
the SM, Higgs boson pair production is dominated by $gg$ fusion
mediated via heavy-quark loops, while the contribution of $q\bar
q$ annihilation is greatly suppressed by the absence of the $HHZ$
coupling and the smallness of the $Hq\bar q$ ($q=u,d,s,c,b$)
couplings. In the MSSM, $gg$ fusion for the pair production of
neutral Higgs bosons can be mediated via both quark loops
\cite{hpair0,hpair1} and squark loops \cite{hpair11,hpair2}, and
the existence of $h^0A^0Z$ and $H^0A^0Z$ couplings at the tree
level leads to $h^0A^0$ and $H^0A^0$ associated productions
through $q\bar q$ annihilations (Drell-Yan-like processes)
\cite{hpair0}. Moreover, since the $\phi b \bar b$ couplings can
be greatly enhanced by large values of $\tan\beta$, there are
potentially important contributions arising from $b \bar b$
annihilation to pair production of neutral Higgs bosons, which
have been studied at the leading-order (LO) \cite{hpair2}.
However, the LO predictions generally have a large uncertainty due
to scale and PDF choices. In this paper, we present the complete
next-to-leading order (NLO) QCD (including SUSY-QCD) calculation
for the cross sections for pair production of neutral Higgs boson
through $b\bar{b}$ annihilation at the LHC. Similar to single
Higgs boson production, for the inclusive production the use of
$b$ quark parton distributions at the LO will improve the
convergence of the perturbative expansion. For simplicity, we
neglect the bottom quark mass except in the Yukawa couplings,
which is valid in all diagrams where the bottom quark is an
initial state parton, according to the simplified
Aivazis-Collins-Olness-Tung (ACOT) scheme \cite{acot}. For
regularization of the ultraviolet (UV), soft and collinear
divergences, both the dimensional regularization (DREG) approach
\cite{DREG} (with naive $\gamma_5$ \cite{gamma5}) and the
dimensional reduction (DRED) scheme \cite{DRED} are used in our
calculations providing a cross check.

This paper is organized as follows. In Sect.II we show the
analytic results for the LO cross sections proceeding through $b
\bar b$ annihilation. In Sect.III we present the details of the
calculations of both the virtual and real parts of the NLO QCD
corrections, and compare the results using DREG with those using
DRED. In Sect.IV  we give the numerical predictions for inclusive
and differential cross sections at the LHC. The relevant coupling
constants and the lengthy analytic expressions are summarized in
Appendices A, B and C.

\section{Leading Order Pair Production of Neutral Higgs Bosons}

\begin{figure}[h!]
\centerline{\epsfig{file=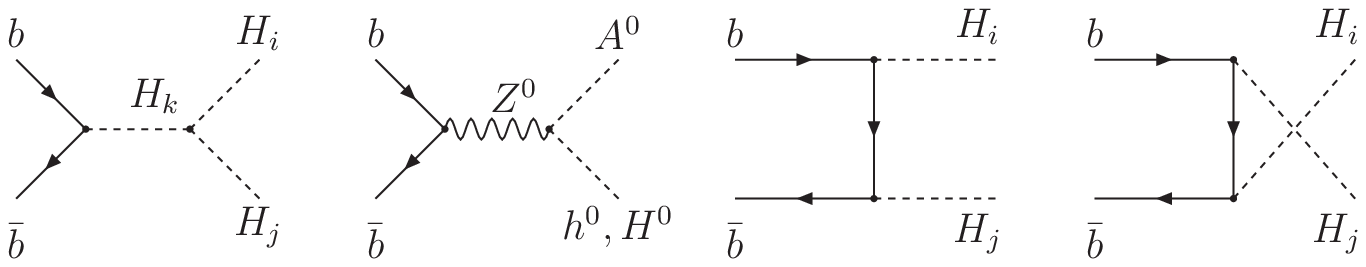, width=400pt}}
\caption[]{Leading order Feynman diagrams for $b\bar{b}\rightarrow
H_iH_j$. \label{fig:feytree}}
\end{figure}

The tree-level Feynman diagrams for the subprocess $b(p_a) \bar
b(p_b) \rightarrow H_i(p_1) H_j(p_2)$, where $H_{i=1,2,3}=H^0,
h^0, A^0$, are shown in Fig.~\ref{fig:feytree}, and its LO
amplitude in $n=4-2\epsilon$ dimensions is
\begin{eqnarray}
M^{B}_{ij}=\mu_r^{2\epsilon} [M_{ij}^{(s)} +M_{ij}^{(t)}
+M_{ij}^{(u)}],
\end{eqnarray}
with
\begin{eqnarray}
&& M_{ij}^{(s)}= -\sum_{k=1}^4\frac{ig^2Y_bm_Z
C_{kij}}{2c_Ws_{H_k}} \bar v(p_b) (a_k P_L + a_k^\ast P_R) u(p_a)
\nonumber
\\
&& \hspace{1.4cm} -\frac{\delta_{i3}g^2Z_j^H}{2c_W^2(s -m_Z^2)}
\bar v(p_b)(\not{\!p}_1 -\not{\!p}_2) (C_{bL} P_L + C_{bR} P_R)
u(p_a), \nonumber
\\
&& M_{ij}^{(t)}=\frac{ig^2 Y_b^2}{t} \bar v(p_b) (a_j a_i^\ast P_L
+ a_i a_j^\ast P_R) \not{\! p}_1 u(p_a), \nonumber
\\
&& M_{ij}^{(u)}=\frac{ig^2 Y_b^2}{u} \bar v(p_b) (a_i a_j^\ast P_L
+ a_j a_i^\ast P_R) \not{\! p}_2 u(p_a),
\end{eqnarray}
where $H_4=G^0$, $c_W\equiv \cos\theta_W$, $Y_b\equiv
m_b/(\sqrt{2}m_W\cos\beta)$, $P_{L,R}\equiv (1\mp \gamma_5)/2$,
$C_{bL}=-1/2+\sin^2\theta_W/3$, $C_{bR}=\sin^2\theta_W/3$, and
$\mu_r$ is a mass parameter introduced to keep the coupling
constant $g$ dimensionless. $s_{X}\equiv s-m_{X}^2+im_{X}\Gamma_X$
is the denominator of the propagator of particle $X$ with mass
$m_X$ and total decay width $\Gamma_X$. $C_{kij}$, $a_k$ and
$Z_j^H$ denote the coefficients appearing in the $H_kH_iH_j$,
$H_kb\bar b$ and $Z^0A^0H_j$ couplings, respectively, and their
explicit expressions are shown in Appendix A. Mandelstam variables
$s$, $t$ and $u$ are defined as follows
\begin{eqnarray}
s=(p_a+p_b)^2, \ \ \ t=(p_a-p_1)^2, \ \ \ u=(p_a-p_2)^2.
\end{eqnarray}
The above amplitude and all of the other calculations in this
paper are carried out in t'Hooft-Feynman gauge.

After the $n$-dimensional phase space integration, the LO parton
level differential cross sections are
\begin{eqnarray}
&& \frac{d^2 \hat{\sigma}_{ij}^B}{dt' du'} =
\frac{1}{1+\delta_{ij}} \frac{\pi
S_{\epsilon}}{s^2\Gamma(1-\epsilon)} (\frac{t'u' - s
m_{H_i}^2}{\mu_r^2s})^{-\epsilon} \Theta(t'u' -s m_{H_i}^2)
\Theta[s- (m_{H_i} +m_{H_j})^2] \nonumber \\
&& \hspace{1.8cm} \times \delta (s+ t +u - m_{H_i}^2 -m_{H_j}^2)
\overline{|M_{ij}^B|}^2
\end{eqnarray}
where $S_{\epsilon} = (4\pi)^{-2 +\epsilon}$, $t'=t-m_{H_i}^2$,
$u'=u-m_{H_i}^2$, the factor $1/(1+\delta_{ij})$ accounts for
identical-particle symmetrization when $H_i=H_j$.
$\overline{|M^B_{ij}|}^2$ is the LO amplitude squared, where the
colors and spins of the out going particles have been summed over,
and the colors and spins of the incoming ones have been averaged
over. The explicit expression for $\overline{|M^B_{ij}|}^2$ is
\begin{eqnarray}
\overline{|M^{B}_{ij}|}^2
    &=& \frac{g^4Y_b^2m_Z^2} {4c_W^2} \sum_{k,l=1}^4 (a_k a_l^\ast
        + a_k^\ast a_l) C_{kij} C_{lij}\frac{s} {s_{H_k} s_{H_l}^\ast}
        + g^4 (t u - m_{H_i}^2 m_{H_j}^2) \bigg \{ \frac{2Y_b^4|a_i
        a_j^\ast t - a_j a_i^\ast u|^2}
        {t^2 u^2}
\nonumber \\
    &+& \delta_{i3} \bigg [ \frac{(Z_j^H)^2(C_{bL}^2 + C_{bR}^2)} {c_W^4
        (s-m_Z)^2} + \frac{2iY_b^2 Z_j^H}{c_W^2 (s - m_Z^2)} (\frac{C_{bR} a_j
        a_i^\ast + C_{bL} a_i a_j^\ast}{t} + \frac{C_{bL} a_j a_i^\ast +
        C_{bR} a_i a_j^\ast}{u}) \bigg ] \bigg \}.
\nonumber \\
\end{eqnarray}

The LO total cross section at the LHC is obtained by convoluting
the parton level cross section with the parton distribution
functions (PDFs) $G_{b,\bar b/p}$ for the proton:
\begin{equation}
\sigma^B_{ij}=\int dx_1dx_2 [G_{b/p}(x_1,\mu_f)G_{\bar
b/p}(x_2,\mu_f)+ (x_1\leftrightarrow x_2)]\hat{\sigma}^{B}_{ij},
\end{equation}
where $\mu_f$ is the factorization scale.

\section{Next-to-Leading order calculations}
The NLO corrections to pair production of neutral Higgs bosons
through $b \bar b$ annihilation consist of the virtual
corrections, generated by loop diagrams of colored particles, and
the real corrections with the radiation of a real gluon or a
massless (anti)bottom quark. For both virtual and real
corrections, we will first present the results in the DREG scheme,
and then in the DRED scheme and compare them.

\subsection{virtual corrections}
\begin{figure}[h!]
\centerline{\epsfig{file=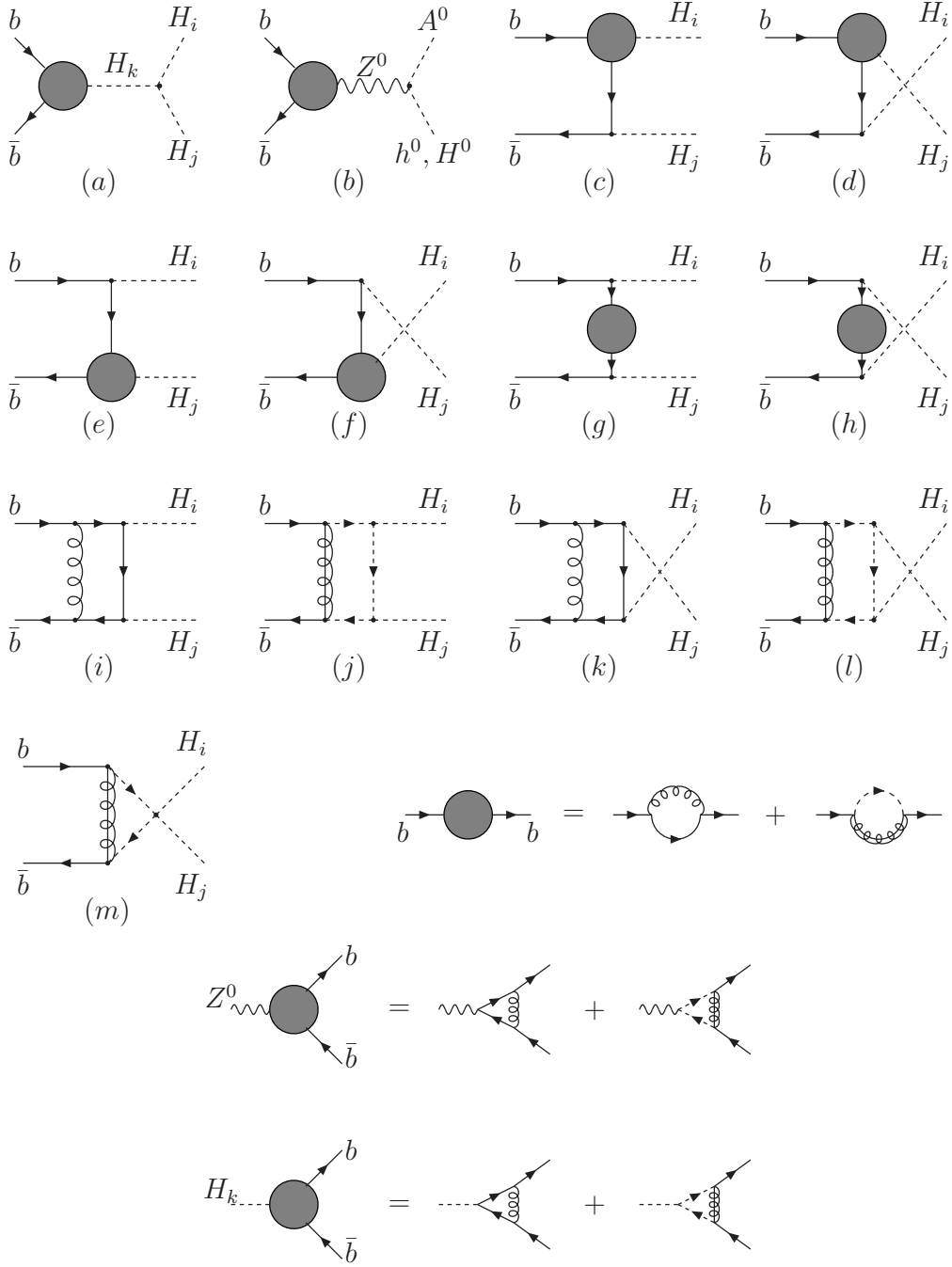, width=380pt}}
\caption[]{Virtual one-loop Feynman diagrams for
$b\bar{b}\rightarrow H_iH_j$. \label{fig:feyvirt}}
\end{figure}

The Feynman diagrams for the virtual corrections to
$b\bar{b}\rightarrow H_iH_j$ are shown in Fig.~\ref{fig:feyvirt}.
In order to remove the UV divergences, we renormalize the bottom
quark mass in the Yukawa couplings and the wave function of the
bottom quark, adopting the on-shell renormalization scheme
\cite{onmass}. The relations between the bare bottom quark mass
$m_{b0}$, the bare wave function $\psi_{b0}$ and their relevant
renormalization constants $\delta m_b$, $\delta Z_{bL(R)}$ are
defined as
\begin{eqnarray}
&& m_{b0}=m_b +\delta m_b, \nonumber
\\
&& \psi_{b0}=(1+\delta Z_{bL})^{\frac{1}{2}}\psi_{bL}+(1+\delta
Z_{bR})^{\frac{1}{2}}\psi_{bR}.
\end{eqnarray}
Calculating the self-energy diagrams in Fig.~\ref{fig:feyvirt}, we
obtain the explicit expressions for $\delta m_b$ and $\delta
Z_{bL(R)}$:
\begin{eqnarray}
&& \frac{\delta m_b}{m_b} =  - \frac{\alpha_s}{4\pi} C_F
\bigg\{3B_0(m_b^2,0,m_b^2) -2 + \sum_{i=1}^2 \bigg[B_1
-\frac{m_{\tilde{g}}}{m_q} \sin2\theta_{\tilde{q}} (-1)^i
B_0\bigg](m_b^2,m_{\tilde{g}}^2,m_{\tilde{b}_i}^2)\bigg\},\nonumber
\\
&& \delta Z_{bL}= \frac{\alpha_s}{2\pi}C_F\sum_{i=1}^2(R^{\tilde
b}_{i1})^2 B_1(0,m_{\tilde{g}}^2,m_{\tilde{b}_i}^2),\nonumber
\\
&& \delta Z_{bR}= \frac{\alpha_s}{2\pi}C_F\sum_{i=1}^2(R^{\tilde
b}_{i2})^2 B_1(0,m_{\tilde{g}}^2,m_{\tilde{b}_i}^2),\nonumber
\end{eqnarray}
where $C_F=4/3$, $B_{0,1}$ are the two-point integrals
\cite{denner}, $m_{\tilde{b}_{1,2}}$ are the sbottom masses,
$m_{\tilde{g}}$ is the gluino mass, and $R^{\tilde b}$ is a $2
\times 2$ matrix defined to rotate the sbottom current eigenstates
into the mass eigenstates:
\begin{equation}
\left(\begin{array}{c} \tilde{b}_1 \\ \tilde{b}_2 \end{array}
\right)= R^{\tilde{b}}\left(\begin{array}{c} \tilde{b}_L \\
\tilde{b}_R \end{array} \right), \ \ \ \ \
R^{\tilde{b}}=\left(\begin{array}{cc} \cos\theta_{\tilde{b}} &
\sin\theta_{\tilde{b}} \\ -\sin\theta_{\tilde{b}} &
\cos\theta_{\tilde{b}}
\end{array} \right)
\end{equation}
with $0 \leq \theta_{\tilde{b}} < \pi$ by convention.
Correspondingly, the mass eigenvalues $m_{\tilde{b}_1}$ and
$m_{\tilde{b}_2}$ (with $m_{\tilde{b}_1}\leq m_{\tilde{b}_2}$) are
given by
\begin{eqnarray}\label{Mq2}
\left(\begin{array}{cc} m_{\tilde{b}_1}^2 & 0 \\ 0 &
m_{\tilde{b}_2}^2 \end{array} \right)=R^{\tilde{b}}
M_{\tilde{b}}^2 (R^{\tilde{b}})^\dag, \ \ \ \ \
M_{\tilde{b}}^2=\left(\begin{array}{cc} m_{\tilde{b}_L}^2 & a_bm_b
\\ a_bm_b & m_{\tilde{b}_R}^2 \end{array} \right)
\end{eqnarray}
with
\begin{eqnarray}
m^2_{\tilde{b}_L} &=& M^2_{\tilde{Q}} +m_b^2 +m_Z^2\cos2\beta
C_{bL}, \nonumber
\\
m^2_{\tilde{b}_R} &=& M^2_{\tilde{D}} +m_b^2 -m_Z^2\cos2\beta
C_{bR}, \nonumber
\\
a_b &=& A_b -\mu\tan\beta.
\end{eqnarray}
Here $M_{\tilde{b}}^2$ is the sbottom mass matrix.
$M_{\tilde{Q},\tilde{D}}$ and $A_{b}$ are soft SUSY-breaking
parameters and $\mu$ is the higgsino mass parameter.

The renormalized virtual amplitudes can be written as
\begin{eqnarray}
M^{V}_{ij}=M^{unren}_{ij}+M^{con}_{ij}.
\end{eqnarray}
Here $M^{unren}_{ij}$ contains the self-energy, vertex and box
corrections, and can be written as
\begin{eqnarray}
M^{unren}_{ij}=\sum_{\alpha=a}^m \frac{iC_F}{16\pi^2} g^2g_s^2
\bar{v}(p_b)[f^\alpha_1 P_L + f^\alpha_2 P_R + \not \! p_1
(f^\alpha_3 P_L +f^\alpha_4 P_R)]u(p_a), \label{formf}
\end{eqnarray}
where $\alpha$ denotes the corresponding diagram in
Fig.~\ref{fig:feyvirt}, and $f_l^\alpha$ $(l=1,2,3,4)$ are the
form factors given explicitly in Appendix B. $M^{con}_{ij}$ is the
corresponding counterterm, and can be separated into
$M^{con(s)}_{ij}$, $M^{con(t)}_{ij}$ and $M^{con(u)}_{ij}$, i.e.
the counterterms for s, t and u channels, respectively:
\begin{eqnarray}
M^{con}_{ij}= M^{con(s)}_{ij} + M^{con(t)}_{ij}+M^{con(u)}_{ij}
\end{eqnarray}
with
\begin{eqnarray}
&& M^{con(s)}_{ij}= -\frac{ig^2Y_b m_Z}{2c_W} [\frac{\delta
m_b}{m_b} +\frac{1}{2}(\delta Z_{bL} +\delta Z_{bR})] \sum_{k=1}^4
\frac{C_{kij}}{s_{H_k}} \bar v (p_b) (a_k P_L + a_k^\ast P_R)
u(p_a) \nonumber
\\
&& \hspace{1.8cm} -\frac{\delta_{i3}g^2Z_j^H}{2c_W^2(s - m_Z^2)}
\bar v (p_b)(\not{\!p}_1 -\not{\!p}_2) (C_{bL} \delta Z_{bL}P_L +
C_{bR} \delta Z_{bR}P_R) u(p_a), \nonumber
\\
&& M^{con(t)}_{ij}= \frac{ig^2Y_b^2}{t} \bar v(p_b) \not{\!p}_1
[a_i a_j^\ast (2\frac{\delta m_b}{m_b} + \delta Z_{bL}) P_L + a_j
a_i^\ast (2\frac{\delta m_b}{m_b} + \delta Z_{bR}) P_R] u(p_a),
\nonumber
\\
&& M^{con(u)}_{ij}= \frac{ig^2Y_b^2}{u} \bar v(p_b) \not{\!p}_2
[a_j a_i^\ast (2\frac{\delta m_b}{m_b} + \delta Z_{bL}) P_L + a_i
a_j^\ast (2\frac{\delta m_b}{m_b} + \delta Z_{bR}) P_R] u(p_a).
\end{eqnarray}

The ${\cal O} (\alpha_s)$ virtual corrections to the differential
cross section can be expressed as
\begin{eqnarray}
&& \frac{d^2 \hat{\sigma}^V_{ij}}{dt' du'} =
\frac{1}{1+\delta_{ij}} \frac{\pi
S_{\epsilon}}{s^2\Gamma(1-\epsilon)}\bigg (\frac{t'u' - s
m_{H_i}^2}{\mu_r^2s}\bigg)^{-\epsilon} \Theta(t'u' -s m_{H_i}^2)
\Theta[s- (m_{H_i} +m_{H_j})^2] \nonumber \\
&& \hspace{1.8cm} \times \delta (s+ t +u - m_{H_i}^2 -m_{H_j}^2) \
2 \ {\rm Re} \overline{(M^V_{ij} M^{B\ast}_{ij})},
\end{eqnarray}
where the renormalized amplitude $M^V_{ij}$ is UV finite, but
still contains the infrared (IR) divergences, and is given by
\begin{eqnarray}
M^V_{ij}|_{IR} =\frac{\alpha_s}{2\pi}
\frac{\Gamma(1-\epsilon)}{1-2\epsilon}
\bigg(\frac{4\pi\mu_r^2}{s}\bigg)^\epsilon
\bigg(\frac{A_2^V}{\epsilon^2}
+\frac{A_1^V}{\epsilon}\bigg)M^B_{ij},
\end{eqnarray}
with
\begin{eqnarray}
A_2^V=-C_F,\qquad  A_1^V=-\frac{3}{2}C_F.
\end{eqnarray}
The coefficients $A_2^V$ and $A_1^V$ are constants, and similar to
those in the pure Drell-Yan-like processes (without color
particles in the final states). These IR divergences include the
soft divergences and the collinear divergences. The soft
divergences will be cancelled after adding the real corrections,
and the remaining collinear divergences can be absorbed into the
redefinition of PDFs \cite{altarelli}, which will be discussed in
the following subsections.

When recalculating the above virtual corrections in the DRED
scheme, one finds that $\delta Z_{bL}$ and $\delta Z_{bR}$ remain
unchanged, however, $\delta m_b$ and the form factors have shifts
which are, respectively, given by
\begin{eqnarray}
\bigg(\frac{\delta m_b}{m_b}\bigg)_{DREG}= \bigg(\frac{\delta
m_b}{m_b}\bigg)_{DRED}+\frac{\alpha_s}{4\pi} C_F,
\end{eqnarray}
and
\begin{eqnarray}
&& \sum_{\alpha=a}^m {f_1^{\alpha}}_{DREG}  =  \sum_{\alpha=a}^m
{f_1^{\alpha}}_{DRED} +\sum_{k=1}^4 \frac{m_ZY_b a_k C_{kij}} {c_W
s_{H_k}}, \nonumber
\\
&& \sum_{\alpha=a}^m {f_2^{\alpha}}_{DREG}  =  \sum_{\alpha=a}^m
{f_2^{\alpha}}_{DRED} +\sum_{k=1}^4 \frac{m_ZY_b a_k^\ast C_{kij}}
{c_W s_{H_k}}, \nonumber
\\
&& \sum_{\alpha=a}^m {f_3^{\alpha}}_{DREG}  =  \sum_{\alpha=a}^m
{f_3^{\alpha}}_{DRED} - \frac{\delta_{i3} iZ_j^H C_{bL}} {c_W^2 (s
-m_Z^2)} - \frac{3Y_b^2}{tu}(u a_ia_j^\ast -t a_j a_i^\ast),
\nonumber
\\
&& \sum_{\alpha=a}^m {f_4^{\alpha}}_{DREG}  =  \sum_{\alpha=a}^m
{f_4^{\alpha}}_{DRED} - \frac{\delta_{i3} iZ_j^H C_{bR}} {c_W^2 (s
-m_Z^2)} - \frac{3Y_b^2}{tu}(u a_ja_i^\ast -t a_i a_j^\ast).
\end{eqnarray}
Thus it is easy to obtain the following relations:
\begin{eqnarray}
&&{M^V_{ij}}_{DREG}={M^V_{ij}}_{DRED}-\frac{\alpha_s}{4\pi}
 C_FM^B_{ij},\\
&&{\sigma^V_{ij}}_{DREG}={\sigma^V_{ij}}_{DRED}-\frac{\alpha_s}{2\pi}
 C_F\sigma^B_{ij}+{\cal O}(\alpha^2_s),
\label{0}
\end{eqnarray}
where $M^B_{ij}$ and $\sigma^B_{ij}$ are independent of the choice
of schemes.

\subsection{Real gluon emission}

\begin{figure}[h!]
\centerline{\epsfig{file=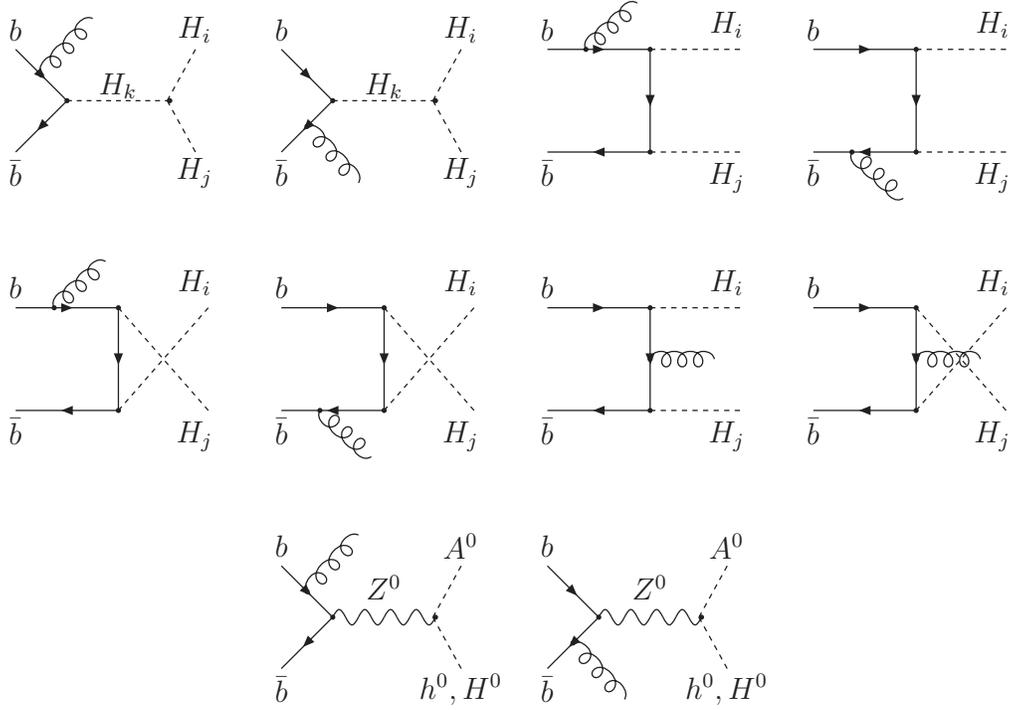, width=380pt}}
\caption[]{Feynman diagrams for the real gluon emission
contributions. \label{fig:feyreal}}
\end{figure}

The feynman diagrams for the real gluon emission process
$b(p_a)\bar{b}(p_b)\rightarrow H_i(p_1)H_j(p_2)+g(p_3)$ are shown
in Fig.~\ref{fig:feyreal}.

The phase space integration for the real gluon emission will
produce soft and collinear singularities, which can be
conveniently isolated by slicing the phase space into different
regions using suitable cut-offs. In this paper, we use the two
cut-off phase space slicing method \cite{cutoff}, which introduces
two arbitrary small cut-offs, i.e. soft cut-off $\delta_s$ and
collinear one $\delta_c$, to decompose the three-body phase space
into three regions.

First, the phase space is separated into two regions by the soft
cut-off $\delta_s$, according to whether the energy of the emitted
gluon is soft, i.e. $E_3\leq \delta_s\sqrt{s}/2$, or hard, i.e.
$E_3> \delta_s\sqrt{s}/2$. Correspondingly, the parton level real
cross section $\hat{\sigma}^{R}_{ij}$ can be written as
\begin{eqnarray}
\hat{\sigma}^{R}_{ij}= \hat{\sigma}^{S}_{ij}
+\hat{\sigma}^{H}_{ij},
\end{eqnarray}
where $\hat{\sigma}^{S}_{ij}$ and $\hat{\sigma}^{H}_{ij}$ are the
contributions from the soft and hard regions, respectively.
$\hat{\sigma}^{S}$ contains all the soft divergences, which can
explicitly be obtained after the integration over the phase space
of the emitted gluon. Next, in order to isolate the remaining
collinear divergences from $\hat{\sigma}^{H}$, the collinear
cut-off $\delta_c$ is introduced to further split the hard gluon
phase space into two regions, according to whether the Mandelstam
variables $u_{1,2}\equiv (p_{a,b}-p_3)^2$ satisfy the collinear
condition $-\delta_c s< u_{1,2}< 0$ or not. We then have
\begin{eqnarray}
\hat{\sigma}^{H}_{ij}= \hat{\sigma}^{HC}_{ij}+
\hat{\sigma}^{\overline{HC}}_{ij},
\end{eqnarray}
where the hard collinear part $\hat{\sigma}^{HC}_{ij}$ contains
the collinear divergences, which also can explicitly be obtained
after the integration over the phase space of the emitted gluon.
And the hard non-collinear part $\hat{\sigma}^{\overline{HC}}$ is
finite and can be numerically computed using standard Monte-Carlo
integration techniques \cite{Monte}, and can be written in the
form
\begin{eqnarray}
d\hat{\sigma}^{\overline{HC}}_{ij}=\frac{1}{2s}
\overline{|M^{b\bar{b}}_{ij}|}^2 d\overline{\Gamma}_3.
\label{nonHC}
\end{eqnarray}
Here $d\overline{\Gamma}_3$ is the hard non-collinear region of
the three-body phase space, and the explicit expressions for
$\overline{|M^{b\bar{b}}_{ij}|}^2$ are given in Appendix C.

In the next two subsections, we will discuss in detail the soft
and hard collinear gluon emission.

\subsubsection{Soft gluon emission}
In the limit that the energy of the emitted gluon becomes small,
i.e. $E_3\leq \delta_s\sqrt{s}/2$, the amplitude squared
$\overline{|M^{R}_{ij}|}^2$ can simply be factorized into the Born
amplitude squared times an eikonal factor $\Phi_{eik}$:
\begin{eqnarray}
\overline{|M^{R}_{ij}(b\bar{b}\rightarrow H_iH_j +g)|}^2
\stackrel{soft}{\rightarrow} (4\pi\alpha_s\mu_r^{2\epsilon})
\overline{|M^{B}_{ij}|}^2 \Phi_{eik},
\end{eqnarray}
where the eikonal factor $\Phi_{eik}$ is given by
\begin{eqnarray}
\Phi_{eik}= C_F\frac{s}{(p_a\cdot p_3)(p_b\cdot p_3)} .
\end{eqnarray}
Moreover, the phase space in the soft limit can also be
factorized:
\begin{eqnarray}
d\Gamma_3(b\bar{b}\rightarrow H_iH_j +g)
\stackrel{soft}{\rightarrow} d\Gamma_2(b\bar{b}\rightarrow H_iH_j
) dS,
\end{eqnarray} Here $dS$ is the integration over the phase space of the soft
gluon, and is given by \cite{cutoff}
\begin{eqnarray}
dS =\frac{1}{2(2\pi)^{3- 2\epsilon}} \int_0^{\delta_s \sqrt{s}/2}
dE_3 E_3^{1 -2\epsilon} d \Omega_{2-2 \epsilon}.
\end{eqnarray}
The parton level cross section in the soft region can then be
expressed as
\begin{eqnarray}\label{soft}
&&\hat{\sigma}^S_{ij} =(4\pi\alpha_s\mu_r^{2\epsilon})\int
d\Gamma_2\overline{|M^{B}_{ij}|}^2 \int dS \Phi_{eik}.
\end{eqnarray}
Using the approach of Ref.~\cite{cutoff}, after the integration
over the soft gluon phase space, Eq.~(\ref{soft}) becomes
\begin{eqnarray}
&&\hat{\sigma}^S_{ij} =\hat{\sigma}^B_{ij} \left
[\frac{\alpha_s}{2\pi}
\frac{\Gamma(1-\epsilon)}{\Gamma(1-2\epsilon)}
\bigg(\frac{4\pi\mu_r^2}{s}\bigg)^\epsilon \right]
\bigg(\frac{A_2^s}{\epsilon^2} +\frac{A_1^s}{\epsilon}
+A_0^s\bigg)
\end{eqnarray}
with
\begin{eqnarray}
A_2^s=2C_F,\qquad A_1^s= -4C_F\ln\delta_s, \qquad
A_0^s=4C_F\ln^2\delta_s.
\end{eqnarray}
These coefficients are the same as the ones in pure Drell-Yan-like
processes, as expected.

\subsubsection{Hard collinear gluon emission}
In the hard collinear region, $E_3> \delta_s\sqrt{s}/2$ and
$-\delta_c s< u_{1,2} < 0$, the emitted hard gluon is collinear to
one of the incoming partons. As a consequence of the factorization
theorems \cite{factor1}, the amplitude squared for
$b\bar{b}\rightarrow H_iH_j +g$ can be factorized into the product
of the Born amplitude squared and the Altarelli-Parisi splitting
function for $b(\bar{b})\rightarrow b(\bar{b})g$
 \cite{altarelli1,factor2}.
\begin{eqnarray}
\overline{|M^{R}_{ij}(b\bar{b}\rightarrow H_iH_j +g)|}^2
\stackrel{collinear}{\rightarrow} (4\pi\alpha_s \mu_r^{2\epsilon})
\overline{|M^{B}_{ij}|}^2 \bigg(\frac{-2P_{bb}(z,\epsilon)}{zu_1}
+\frac{-2P_{\bar{b}\bar{b}}(z,\epsilon)}{zu_2}\bigg),
\end{eqnarray}
Here $z$ denotes the fraction of incoming parton $b(\bar{b})$'s
momentum carried by parton $b(\bar{b})$ with the emitted gluon
taking a fraction $(1-z)$, and $P_{ij}(z,\epsilon)$ are the
unregulated splitting functions in $n=4-2\epsilon$ dimensions for
$0<z<1$, which can be related to the usual Altarelli-Parisi
splitting kernels \cite{altarelli1} as follows:
$P_{ij}(z,\epsilon)=P_{ij}(z) +\epsilon P_{ij}'(z)$, explicitly
\begin{eqnarray}
&& P_{bb}(z)=P_{\bar{b}\bar{b}}(z)=C_F \frac{1
+z^2}{1-z}+C_F\frac{3}{2}\delta(1-z), \\
&&P_{bb}'(z)=P_{\bar{b}\bar{b}}'(z)= -C_F
(1-z)+C_F\frac{1}{2}\delta(1-z).
\end{eqnarray}
Moreover, the three-body phase space can also be factorized in the
collinear limit, and, for example, in the limit $-\delta_c s< u_1
< 0$ it has the following form \cite{cutoff}:
\begin{eqnarray}
d\Gamma_3(b\bar{b}\rightarrow H_iH_j +g)
\stackrel{collinear}{\rightarrow} d\Gamma_2(b\bar{b}\rightarrow
H_iH_j; s'=zs) \frac{(4\pi)^\epsilon}{16\pi^2\Gamma(1- \epsilon)}
dz du_1[-(1 -z)u_1]^{-\epsilon}.
\end{eqnarray}
Here the two-body phase space is evaluated at a squared
parton-parton energy of $zs$. Thus the three-body cross section in
the hard collinear region is given by \cite{cutoff}
\begin{eqnarray}
&& d\sigma^{HC}_{ij} =d\hat{\sigma}^B_{ij}
\bigg[\frac{\alpha_s}{2\pi} \frac{\Gamma(1-\epsilon)}
{\Gamma(1-2\epsilon)} (\frac{4\pi\mu_r^2}{s})^\epsilon\bigg]
(-\frac{1}{\epsilon}) \delta_c^{-\epsilon}
\bigg[P_{bb}(z,\epsilon)G_{b/p}(x_1/z)G_{\bar{b}/p}(x_2) \nonumber
\\ && \hspace{1.4cm} + P_{\bar{b}\bar{b}}
(z,\epsilon)G_{\bar{b}/p}(x_1/z) G_{b/p}(x_2) +(x_1\leftrightarrow
x_2)\bigg] \frac{dz}{z} (\frac{1 -z}{z})^{-\epsilon} dx_1 dx_2,
\end{eqnarray}
where $G_{b(\bar{b})/p}(x)$ is the bare PDF.

\subsection{Massless $b(\bar b)$ emission}
\begin{figure}[h!]
\centerline{\epsfig{file=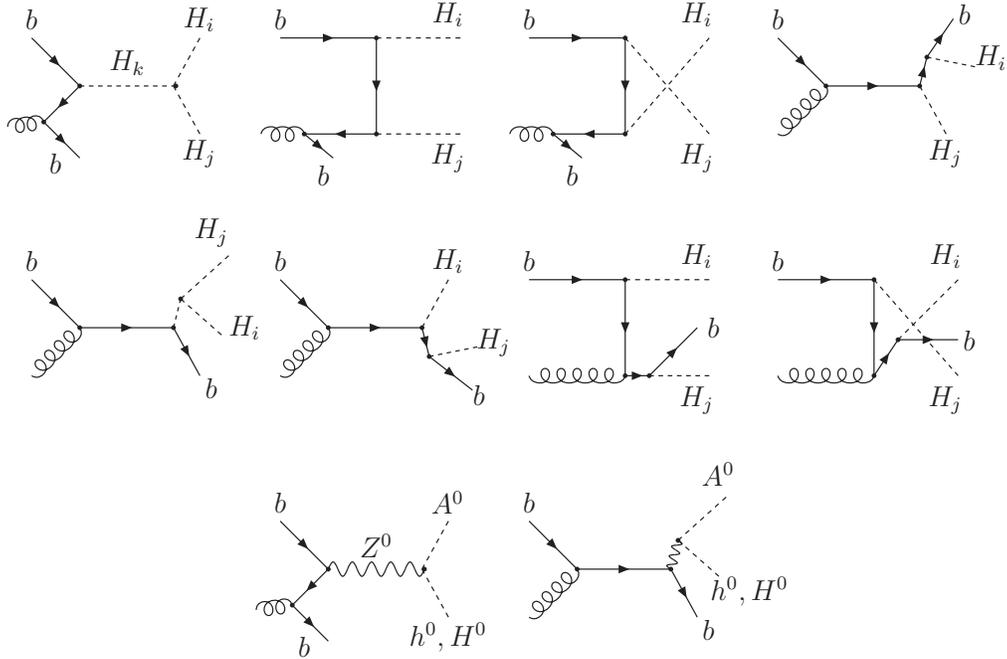, width=380pt}}
\caption[]{Feynman diagrams for the massless bottom quark emission
contributions. \label{fig:feysplit}}
\end{figure}

In addition to real gluon emission, a second set of real emission
corrections to the inclusive cross section for $pp\rightarrow
H_iH_j$ at NLO involves the processes with an additional massless
$b(\bar b)$ in the final state:
\begin{eqnarray}
gb\rightarrow H_iH_jb, \qquad g\bar{b}\rightarrow H_iH_j\bar b.
\nonumber
\end{eqnarray}
The relevant feynman diagrams for massless $b$ emission are shown
in Fig.~\ref{fig:feysplit}, and the diagrams for $\bar b$-emission
are similar and omitted here.

Since the contributions from real massless $b(\bar b)$ emission
contain initial state collinear singularities, we also need to use
the two cut-off phase space slicing method \cite{cutoff} to
isolate these collinear divergences. But we only split the phase
space into two regions, because there are no soft divergences.
Consequently, using the approach in Ref.~\cite{cutoff}, the cross
sections for the processes with an additional massless $b(\bar b)$
in the final state can be expressed as
\begin{eqnarray}\label{accd}
&& d\sigma^{add}_{ij}= \sum_{\alpha=b,\bar{b}}
\hat{\sigma}^{\overline{C}}_{ij}(g\alpha\rightarrow H_iH_j +X)
[G_{g/p}(x_1) G_{\alpha/p}(x_2) +(x_1\leftrightarrow x_2)]
dx_1dx_2 \nonumber
\\&& \hspace{1.0cm}
+d\hat{\sigma}^B_{ij} \bigg[\frac{\alpha_s}{2\pi}
\frac{\Gamma(1-\epsilon)} {\Gamma(1-2\epsilon)}
(\frac{4\pi\mu^2_r}{s})^\epsilon\bigg] (-\frac{1}{\epsilon})
\delta_c^{-\epsilon}
\bigg[P_{bg}(z,\epsilon)G_{g/p}(x_1/z)G_{\bar{b}/p}(x_2) \nonumber
\\ && \hspace{1.0cm}
+G_{b/p}(x_1)P_{\bar{b}g}(z,\epsilon)G_{g/p}(x_2/z)+(x_1\leftrightarrow
x_2)\bigg] \frac{dz}{z} \bigg(\frac{1 -z}{z}\bigg)^{-\epsilon}
dx_1 dx_2,
\end{eqnarray}
where
\begin{eqnarray}
&& P_{bg}(z) =P_{\bar{b}g}(z)=\frac{1}{2}[z^2 +(1-z)^2],
\hspace{2.0cm} P_{bg}'(z)=P_{\bar{b}g}'(z)=-z(1-z).
\end{eqnarray}
The first term in Eq.(\ref{accd}) represents the non-collinear
cross sections for the two processes, which can also be written in
the form ($\alpha=b, \bar b$)
\begin{eqnarray}
d\hat{\sigma}^{\overline{C}}_{ij}=\frac{1}{2s}
\overline{|M^{g\alpha}_{ij}|}^2 d\overline{\Gamma}_3, \label{qHC}
\end{eqnarray}
where $d\overline{\Gamma}_3$ is the three body phase space in the
non-collinear region. The explicit expressions for
$\overline{|M^{g\alpha}_{ij}|}^2$ can be obtained from
$\overline{|M^{b\bar b}_{ij}|}^2$ by crossing symmetry. The second
term in Eq.(\ref{accd}) represents the collinear singular cross
sections.

\subsection{Mass factorization}
As mentioned above, after adding the renormalized virtual
corrections and the real corrections, the parton level cross
sections still contain collinear divergences, which can be
absorbed into the redefinition of the PDFs at NLO, in general
called mass factorization \cite{altarelli}. This procedure in
practice means that first we convolute the partonic cross section
with the bare PDF $G_{\alpha/p}(x)$, and then use the renormalized
PDF $G_{\alpha/p}(x,\mu_f)$ to replace $G_{\alpha/p}(x)$. In the
$\overline{\rm MS}$ convention, the scale dependent PDF
$G_{\alpha/p}(x,\mu_f)$ is given by \cite{cutoff}
\begin{eqnarray}
G_{\alpha/p}(x,\mu_f)= G_{\alpha/p}(x)+
\sum_{\beta}(-\frac{1}{\epsilon})\bigg [\frac{\alpha_s}{2\pi}
\frac{\Gamma(1 -\epsilon)}{\Gamma(1 -2\epsilon)} \bigg(\frac{4\pi
\mu_r^2}{\mu_f^2}\bigg)^\epsilon\bigg]  \int_x^1 \frac{dz}{z}
P_{\alpha\beta} (z) G_{\beta/p}(x/z).
\end{eqnarray}
This replacement will produce a collinear singular counterterm,
which is combined with the hard collinear contributions to result
in, as the definition in Ref.~\cite{cutoff}, the ${\cal O}
(\alpha_s)$ expression for the remaining collinear contribution:
\begin{eqnarray}
&& d\sigma^{coll}_{ij}=d\hat{\sigma}^B_{ij}
\bigg[\frac{\alpha_s}{2\pi} \frac{\Gamma(1-\epsilon)}
{\Gamma(1-2\epsilon)} \bigg(\frac{4\pi\mu^2_r}{s}\bigg)^\epsilon
\bigg] \{\tilde{G}_{b/p}(x_1,\mu_f) G_{\bar{b}/p}(x_2,\mu_f) +
G_{b/p}(x_1,\mu_f) \tilde{G}_{\bar{b}/p}(x_2,\mu_f) \nonumber
\\ && \hspace{1.2cm}
+\sum_{\alpha=b,\bar{b}}\bigg[\frac{A_1^{sc}(\alpha\rightarrow
\alpha g)}{\epsilon} +A_0^{sc}(\alpha\rightarrow \alpha
g)\bigg]G_{b/p}(x_1,\mu_f) G_{\bar{b}/p}(x_2,\mu_f) \nonumber
\\ && \hspace{1.2cm}
+(x_1\leftrightarrow x_2)\} dx_1dx_2,\label{11}
\end{eqnarray}
where
\begin{eqnarray}
&& A_1^{sc}(b\rightarrow bg)=A_1^{sc}(\bar{b}\rightarrow
\bar{b}g)=C_F(2\ln\delta_s
+3/2), \\
&& A_0^{sc}=A_1^{sc}\ln(\frac{s}{\mu_f^2}), \\
&& \tilde{G}_{\alpha/p}(x,\mu_f)=\sum_{\beta}\int_x^{1-
\delta_s\delta_{\alpha\beta}} \frac{dy}{y}
G_{\beta/p}(x/y,\mu_f)\tilde{P}_{\alpha\beta}(y)
\end{eqnarray}
with
\begin{eqnarray}
\tilde{P}_{\alpha\beta}(y)=P_{\alpha\beta} \ln(\delta_c
\frac{1-y}{y} \frac{s}{\mu_f^2}) -P_{\alpha\beta}'(y).
\end{eqnarray}

Finally, the NLO total cross section for $pp\rightarrow H_iH_j$ in
the $\overline{MS}$ factorization scheme is
\begin{eqnarray}
&& \sigma^{NLO}_{ij}= \int \{dx_1dx_2
\bigg[G_{b/p}(x_1,\mu_f)G_{\bar{b}/p}(x_2,\mu_f)+
(x_1\leftrightarrow x_2)\bigg](\hat{\sigma}^{B}_{ij} +
\hat{\sigma}^{V}_{ij}+ \hat{\sigma}^{S}_{ij}
+\hat{\sigma}^{\overline{HC}}_{ij}) +\sigma^{coll}_{ij}\}
\nonumber
\\ && \hspace{0.4cm} +\sum_{\alpha=b,\bar{b}}\int dx_1dx_2
\bigg[G_{g/p}(x_1,\mu_f) G_{\alpha/p}(x_2,\mu_f)
+(x_1\leftrightarrow x_2)\bigg]
\hat{\sigma}^{\overline{C}}_{ij}(g\alpha\rightarrow H_iH_j
+X).\label{t}
\end{eqnarray}
Note that the above expression contains no singularities since
$2A_2^V +A_2^s =0$ and $2A_1^V +A_1^s +A_1^{sc}(b\rightarrow bg)
+A_1^{sc}(\bar{b}\rightarrow \bar{b}g) =0$.

\subsection{Real corrections and NLO total cross sections in the DRED scheme}
Above we gave the real corrections and NLO total cross sections in
the DREG scheme, and next we show the corresponding results in the
DRED scheme, where the contributions from soft gluon emission
remain the same, while those from hard collinear gluon emission
and massless (anti)quark emission are different. These differences
arise from the splitting functions and the PDFs.

First, the splitting functions in the DRED scheme have no
$\epsilon$ parts, and we have
\begin{eqnarray}
P_{ij}(z,\epsilon)_{DRED}=P_{ij}(z).\label{12}
\end{eqnarray}
Then from Eq.~(\ref{11}) and (\ref{12}) we obtain
\begin{eqnarray}
&&{\sigma^{coll}_{ij}}_{DREG}={\sigma^{coll}_{ij}}_{DRED}-\frac{\alpha_s}{2\pi}
\{\sum_{\beta}\int_{x_1}^{1-\delta_s\delta_{b\beta}} \frac{dy}{y}
G_{\beta/p}(x_1/y,\mu_f){P^\prime}_{b\beta}(y)
G_{\bar{b}/p}(x_2,\mu_f)\nonumber
\\&& \hspace{0.2cm}+ \sum_{\beta}\int_{x_2}^{1-\delta_s\delta_{\bar{b}\beta}}
\frac{dy}{y} G_{\beta/p}(x_2/y,\mu_f){P^\prime}_{\bar{b}\beta}(y)
G_{b/p}(x_1,\mu_f) +(x_1\leftrightarrow x_2)\}\hat{\sigma}^B
dx_1dx_2,\label{1}
\end{eqnarray}
Secondly, the PDFs in the DRED and DREG schemes are related
\cite{diffPDF}:
\begin{eqnarray}
G_{\alpha/p}(x,\mu_f)_{DREG}=G_{\alpha/p}(x,\mu_f)_{DRED}+\frac{\alpha_s}{2\pi}\sum_{\beta}
\int_{x}^{1}\frac{dy}{y}{P^\prime}_{\alpha\beta}(x/y).
\end{eqnarray}
Substituting into the formula for the Born cross section, we
obtain an additional difference at the ${\cal O}(\alpha_s)$ level
arising from the PDFs:
\begin{eqnarray}
&&{\sigma^{B}_{ij}}_{DREG}={\sigma^{B}_{ij}}_{DRED}+\frac{\alpha_s}{2\pi}
\{\sum_{\beta}\int_{x_1}^{1} \frac{dy}{y}
G_{\beta/p}(x_1/y,\mu_f)_{DRED}{P^\prime}_{b\beta}(y)
G_{\bar{b}/p}(x_2,\mu_f)_{DRED} \nonumber
\\
&& \hspace{0.8cm}+ \sum_{\beta}\int_{x_2}^{1} \frac{dy}{y}
G_{\beta/p}(x_2/y,\mu_f)_{DRED}{P^\prime}_{\bar{b}\beta}(y)
G_{b/p}(x_1,\mu_f)_{DRED} +(x_1\leftrightarrow
x_2)\}\hat{\sigma}^B dx_1dx_2. \nonumber
\\
\label{2}
\end{eqnarray}
Equations (\ref{1}) and (\ref{2}) are very similar except for the
limits of the integral over y in the two expressions. Substituting
the equations (\ref{1}), (\ref{2}) and (\ref{0}) into (\ref{t}),
we obtain the relation of the NLO total cross sections in the two
schemes:
\begin{eqnarray}
&&{\sigma^{NLO}_{ij}}_{DREG}={\sigma^{NLO}_{ij}}_{DRED}+\frac{\alpha_s}{2\pi}
\{\sum_{\beta}\int_{1-\delta_s\delta_{b\beta}}^{1} \frac{dy}{y}
G_{\beta/p}(x_1/y,\mu_f){P^\prime}_{b\beta}(y)
G_{\bar{b}/p}(x_2,\mu_f)\nonumber
\\&& \hspace{1.2cm}+ \sum_{\beta}\int_{1-\delta_s\delta_{\bar{b}\beta}}^{1}
\frac{dy}{y} G_{\beta/p}(x_2/y,\mu_f){P^\prime}_{\bar{b}\beta}(y)
G_{b/p}(x_1,\mu_f) +(x_1\leftrightarrow x_2)\}\hat{\sigma}^B_{ij}
dx_1dx_2\nonumber
\\&& \hspace{1.2cm}-\frac{\alpha_s}{2\pi}C_F{\sigma^B_{ij}}+{\cal O}
(\alpha^2_s).
\end{eqnarray}
Using the explicit expressions for the $\epsilon$ parts of the
splitting functions $P^\prime$, we find
\begin{eqnarray}
{\sigma^{NLO}_{ij}}_{DREG}={\sigma^{NLO}_{ij}}_{DRED}+{\cal O}
(\alpha^2_s).
\end{eqnarray}
Therefore, the NLO total cross sections in the two schemes are the
same.

\subsection{Differential cross section in transverse momentum and invariant mass}
In this subsection we present the differential cross section in
the transverse momentum $p_T$ and the invariant mass. Using the
notations defined in Ref.~\cite{beenakker2}, the differential
distribution with respect to $p_T$ and $y$ of $H_i$ for the
processes
\begin{eqnarray}
p(p_a)+ p(p_b) \rightarrow H_i (p_1) +H_j (p_2) [+
g(p_3)/b(p_3)/\bar{b}(p_3)]
\end{eqnarray}
is given by
\begin{eqnarray} \label{integralpt}
\frac{d^2\sigma_{ij}}{dp_T dy} =2 p_T S \sum_{\alpha,\beta}
\int_{x_1^-}^1 dx_1 \int_{x_2^-}^1 dx_2 x_1
G_{\alpha/p}(x_1,\mu_f) x_2 G_{\beta/p}(x_2,\mu_f) \frac{d^2
\hat{\sigma}_{ij}}{dt' du'},
\end{eqnarray}
where $\sqrt{S}$ is the total center-of-mass energy of the
collider, and
\begin{eqnarray}
&& p_T^2= \frac{T_2U_2}{S} -m_{H_i}^2, \hspace{1.8cm}
y=\frac{1}{2}\ln(\frac{T_2}{U_2}), \nonumber \\ && x_1^-=
\frac{-T_2 -m_{H_i}^2 +m_{H_j}^2}{S +U_2}, \ \ \ \ \ \ x_2^-=
\frac{-x_1U_2 -m_{H_i}^2 +m_{H_j}^2}{x_1S +T_2}
\end{eqnarray}
with $T_2=(p_b-p_1)^2 - m_{H_i}^2$ and $U_2=(p_a-p_1)^2 -
m_{H_i}^2$. The limits of integration over $y$ and $p_T$ are
\begin{eqnarray}
-y^{max}(p_T)\leq y \leq y^{max}(p_T), \hspace{1.5cm}  0\leq p_T
\leq p_T^{max},
\end{eqnarray}
with
\begin{eqnarray}
&& y^{max}(p_T)={\rm arccosh}\bigg(\frac{S+ m_{H_i}^2-
m_{H_j}^2}{2\sqrt{S (p_T^2 +m_{H_i}^2)}}\bigg), \nonumber
\\ && p_T^{max}=\frac{1}{2\sqrt{S}} \sqrt{(S
+m_{H_i}^2 -m_{H_j}^2)^2 -4m_{H_i}^2S} \ .
\end{eqnarray}

The differential distribution with respect to the invariant mass
$M_{H_iH_j}$ is given by
\begin{eqnarray}
\frac{d\sigma}{dM_{H_iH_j}} = \frac{2M_{H_iH_j}}{S}
\sum_{\alpha,\beta}\frac{d{\cal{L}}_{H_iH_j}^{\alpha\beta}}
{d\tau}\hat{\sigma}_{\alpha\beta}(\tau S), \label{invariant}
\end{eqnarray}
where $d{\cal{L}}_{H_iH_j}^{\alpha\beta}/d\tau$ is the parton
luminosity:
\begin{eqnarray}
\frac{d{\cal{L}}_{H_iH_j}^{\alpha\beta}}{d\tau}=\int^1_\tau
\frac{dx}{x}\bigg[G_{\alpha/p}(x,\mu_f)G_{\beta/p}(\tau/x,\mu_f)
\bigg],
\end{eqnarray}
with
\begin{eqnarray}
&&M_{H_iH_j}\equiv \sqrt{(p_1+p_2)^2}\geq (m_{H_i}+m_{H_j}),\qquad
\tau\equiv M^2_{H_iH_j}/S.
\end{eqnarray}

\section{Numerical results and conclusions}
In this section, we present the numerical results for total and
differential cross sections for pair production of neutral Higgs
bosons at the LHC. In our numerical calculations, the SM
parameters were taken to be $\alpha_{\rm ew}(m_W)=1/128$,
$m_W=80.425$ GeV,  $m_Z=91.1876$ GeV and $m_t=178.1$ GeV
\cite{parameter}. We used the two-loop evaluation for
$\alpha_s(Q)$ \cite{runningalphas} $(\alpha_s(M_Z)=0.118)$ and
CTEQ6M PDFs \cite{CTEQ} throughout the calculations of the NLO
(LO) cross sections unless specified . Moreover, in order to
improve the perturbative calculations, we took the running mass
$m_b(Q)$ evaluated by the NLO formula \cite{runningmb}:
\begin{equation}
m_b(Q)=U_6(Q,m_t)U_5(m_t,m_b)m_b(m_b)
\end{equation}
with $m_b(m_b)=4.25$ GeV \cite{mb}. The evolution factor $U_f$ is
\begin{eqnarray}
U_f(Q_2,Q_1)=\bigg(\frac{\alpha_s(Q_2)}{\alpha_s(Q_1)}\bigg)^{d^{(f)}}
\bigg[1+\frac{\alpha_s(Q_1)-\alpha_s(Q_2)}{4\pi}J^{(f)}\bigg], \nonumber \\
d^{(f)}=\frac{12}{33-2f}, \hspace{1.0cm}
J^{(f)}=-\frac{8982-504f+40f^2}{3(33-2f)^2}.
\end{eqnarray}
In addition, to also improve the perturbation calculations,
especially for large $\tan\beta$, we made the following SUSY
replacements in the tree-level couplings \cite{runningmb}:
\begin{eqnarray}
&& m_b(Q) \ \ \rightarrow \ \ \frac{m_b(Q)}{1+\Delta m_b},
\label{deltamb}
\\
&& \Delta m_b=\frac{2\alpha_s}{3\pi}m_{\tilde{g}}\mu\tan\beta
I(m_{\tilde{b}_1},m_{\tilde{b}_2},m_{\tilde{g}})
+\frac{g^2m_t^2}{32\pi^2 m_W^2 \sin^2\beta}\mu A_t\tan\beta
I(m_{\tilde{t}_1},m_{\tilde{t}_2},\mu) \nonumber \\
&& \hspace{1.0cm} -\frac{g^2}{16\pi^2}\mu M_2\tan\beta
\sum_{i=1}^2 \bigg[(R^{\tilde{t}}_{i1})^2
I(m_{\tilde{t}_i},M_2,\mu) + \frac{1}{2}(R^{\tilde{b}}_{i1})^2
I(m_{\tilde{b}_i},M_2,\mu)\bigg] \label{deltamb1}
\end{eqnarray}
with
\begin{eqnarray}
I(a,b,c)=\frac{1}{(a^2-b^2)(b^2-c^2)(a^2-c^2)}
(a^2b^2\log\frac{a^2}{b^2} +b^2c^2\log\frac{b^2}{c^2}
+c^2a^2\log\frac{c^2}{a^2}).
\end{eqnarray}
It is necessary, to avoid double counting, to subtract these
(SUSY-)QCD corrections from the renormalization constant $\delta
m_b$ in the following numerical calculations.

For the MSSM parameters, we chose $m_{\frac{1}{2}}$, $m_0$, $A_0$,
$\tan\beta$ and the sign of $\mu$ as input parameters, where
$m_{\frac{1}{2}}$, $m_0$ and $A_0$ are the universal gaugino mass
, scalar mass at the GUT scale and the trilinear soft breaking
parameter in the superpotential terms, respectively. Specifically,
we took $m_{\frac{1}{2}}=170$ GeV, $A_0=200$ GeV, and tuned $m_0$
to obtain the desired value of $m_{A^0}$, while $\tan\beta$ and
sign of $\mu$ are free. All other MSSM parameters are determined
in the minimal supergravity (mSUGRA) scenario by the program
package SUSPECT 2.3 \cite{suspect}. In particular, we used running
$\overline{\rm DR}$ Higgs masses at $m_Z$, which include the full
one--loop corrections, as well as the two--loop corrections
controlled by the strong gauge coupling and the Yukawa couplings
of the third generation fermions \cite{higgsmass}. Moreover, since
an $s$-channel resonance can occur in the process $pp\rightarrow
h^0h^0+X$ when $m_{H^0}>2m_{h^0}$, we adopted the program package
HDECAY 3.101 \cite{hdecay} to determine the total decay widths of
the Higgs bosons. For example, in the case of $\tan\beta=40$,
$m_{A^0}=250$ GeV and $\mu<0$, we have $m_{h^0}=108.7$ GeV,
$m_{H^0}=250.3$ GeV, $\Gamma_{h^0}=8$ MeV, $\Gamma_{H^0}=13.1$
GeV, $\Gamma_{A^0}=13.5$ GeV, $m_{\tilde g}=454$ GeV, $m_{\tilde
b_1}=422$ GeV, and $m_{\tilde b_2}=493$ GeV.

For the renormalization and factorization scales, we always chose
$\mu_r=m_{\rm av}\equiv(m_{H_i}+m_{H_j})/2$ and $\mu_f=m_{\rm av}$
unless specified.

\begin{figure}[h!]
\centerline{\epsfig{file=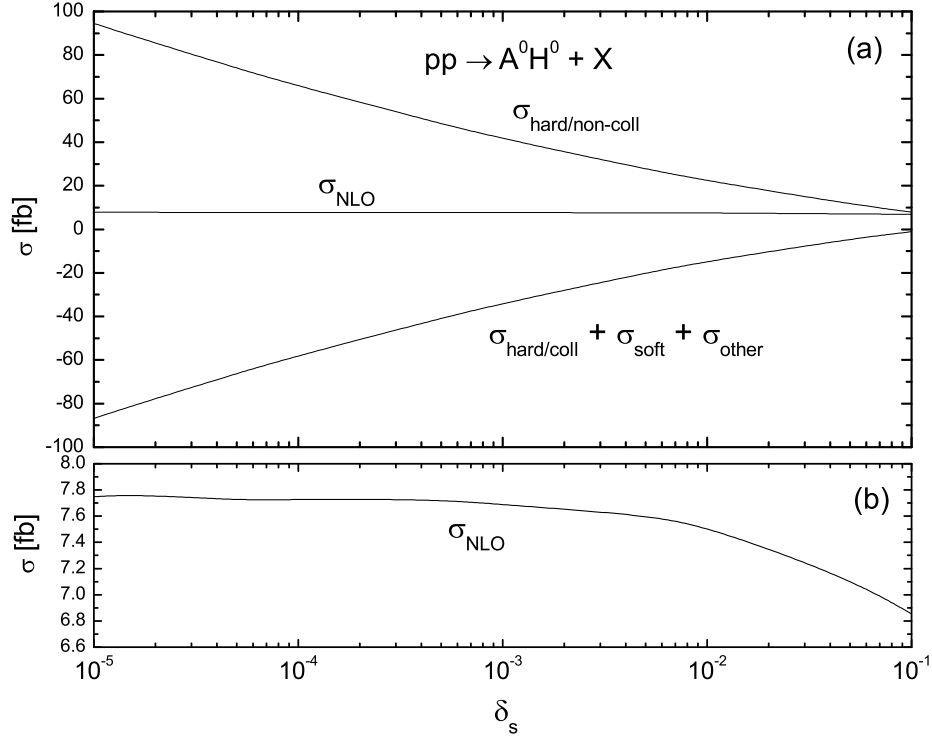,width=350pt}}
\caption[]{Dependence of the total cross sections for $A^0H^0$
production through $b\bar b$ annihilation at the LHC on the
cut-off $\delta_s$, assuming $m_{A^0}=250$ GeV, $\tan\beta=40$,
$m_{1/2}=170$ GeV, $A_0=200$ GeV, $\mu<0$ and
$\delta_c=\delta_s/50$. \label{fig:deltas}}
\end{figure}

In Fig.~\ref{fig:deltas}, we chose $A^0H^0$ production through
$b\bar b$ annihilation as an example to show that it is reasonable
to use the two cut-off phase space slicing method in our NLO QCD
calculations, i.e. the dependence of the NLO QCD predictions on
the arbitrary cut-offs $\delta_s$ and $\delta_c$ is indeed very
weak, as shown in Ref.~\cite{cutoff}. Here $\sigma_{\rm other}$
includes the contributions from the Born cross section and the
virtual corrections, which are cut-off independent. Both the soft
plus hard collinear contributions and the hard non-collinear
contributions depend strongly on the cutoffs and, especially for
the small cut-offs ($\delta_s<10^{-5}$), each is about ten times
larger than the LO total cross section. However, the two
contributions ($\sigma_{soft}+\sigma_{hard/coll}+\sigma_{virtual}$
and $\sigma_{hard/non-coll}$) nearly cancel each other completely,
especially for the cut-off $\delta_s$ between $5\times 10^{-5}$
and $10^{-3}$, where the final results for $\sigma_{NLO}$ are
almost independent of the cut-offs and very near 7.7 fb.
Therefore, we will take $\delta_s=10^{-4}$ and
$\delta_c=\delta_s/50$ in the numerical calculations below.

\begin{figure}[h!]
\centerline{\epsfig{file=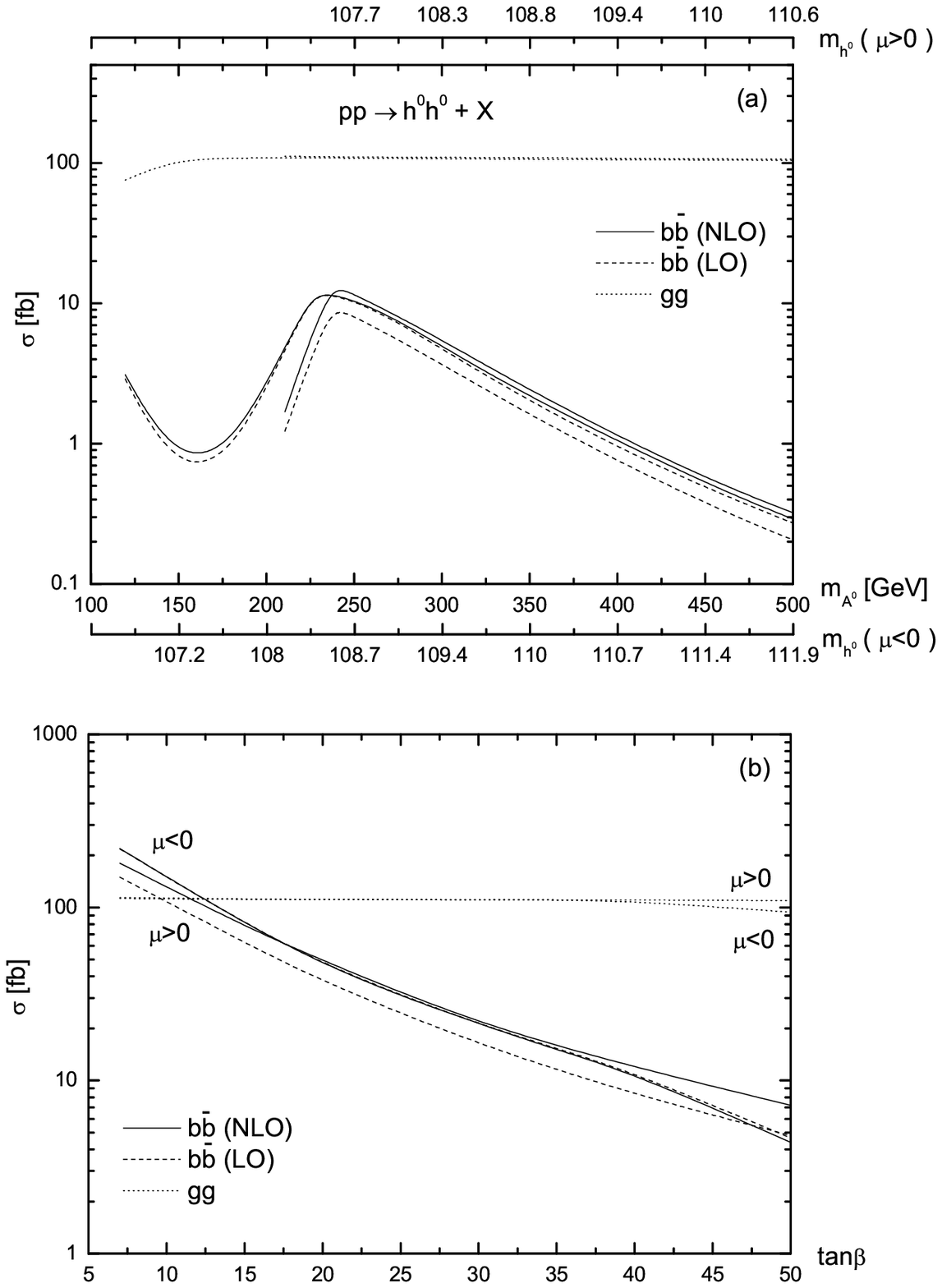,width=350pt}} \caption[]{Total
cross sections for the $h^0h^0$ production at the LHC (a) as
functions of $m_{A^0}$ for $\tan\beta=40$, $\mu<0$ (starting from
$m_{A^0}=120$ GeV) and $\tan\beta=40$, $\mu>0$ (starting from
$m_{A^0}=210$ GeV), and (b) as functions of $\tan\beta$ for
$m_{A^0}=250$ GeV, assuming $m_{1/2}=170$ GeV and $A_0=200$ GeV.
\label{fig:hlhl}}
\end{figure}

\begin{figure}[h!]
\centerline{\epsfig{file=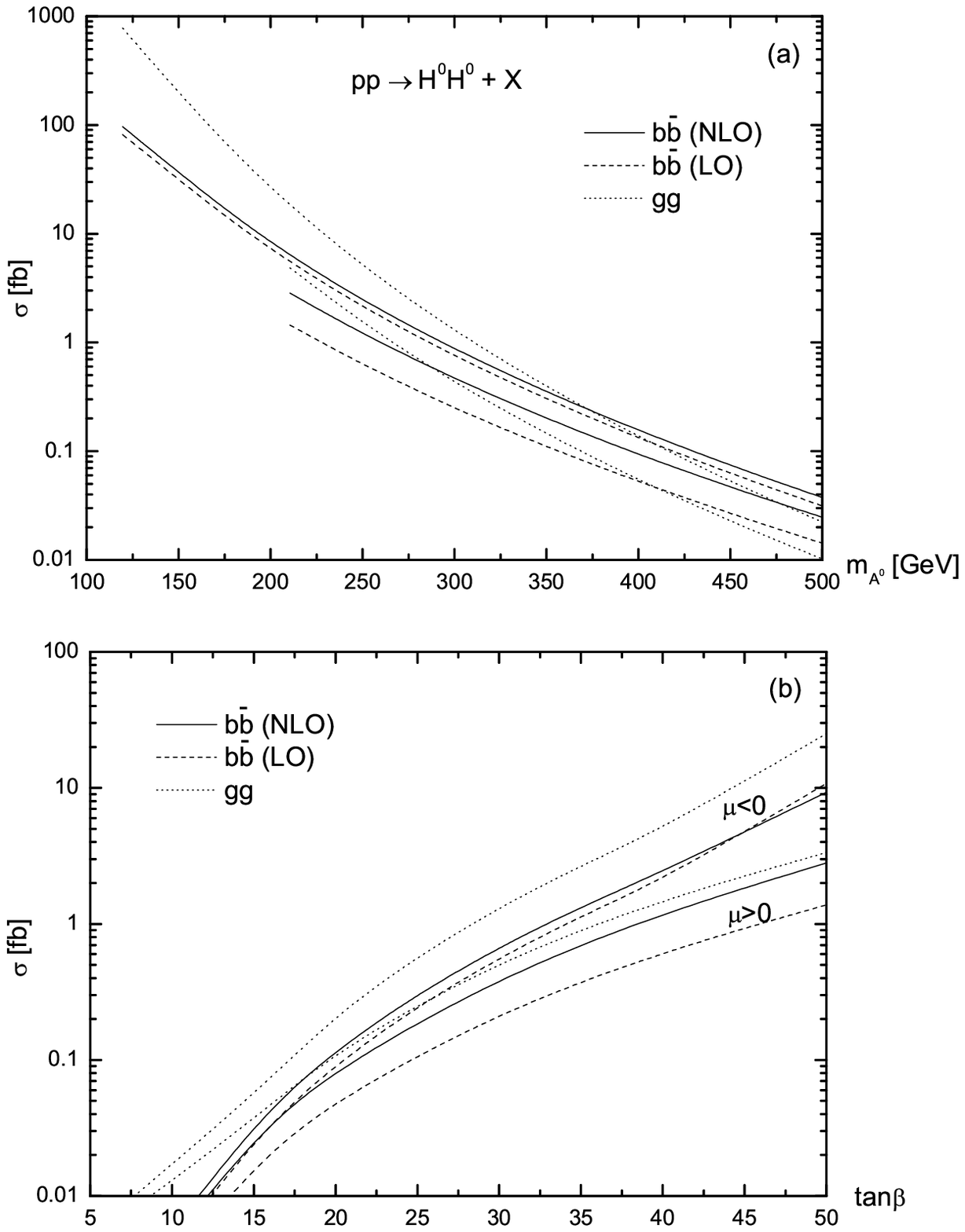,width=350pt}} \caption[]{Total
cross sections for the $H^0H^0$ production at the LHC (a) as
functions of $m_{A^0}$ for $\tan\beta=40$, $\mu<0$ (starting from
$m_{A^0}=120$ GeV) and $\tan\beta=40$, $\mu>0$ (starting from
$m_{A^0}=210$ GeV), and (b) as functions of $\tan\beta$ for
$m_{A^0}=250$ GeV, assuming $m_{1/2}=170$ GeV and $A_0=200$ GeV.
\label{fig:hh}}
\end{figure}

\begin{figure}[h!]
\centerline{\epsfig{file=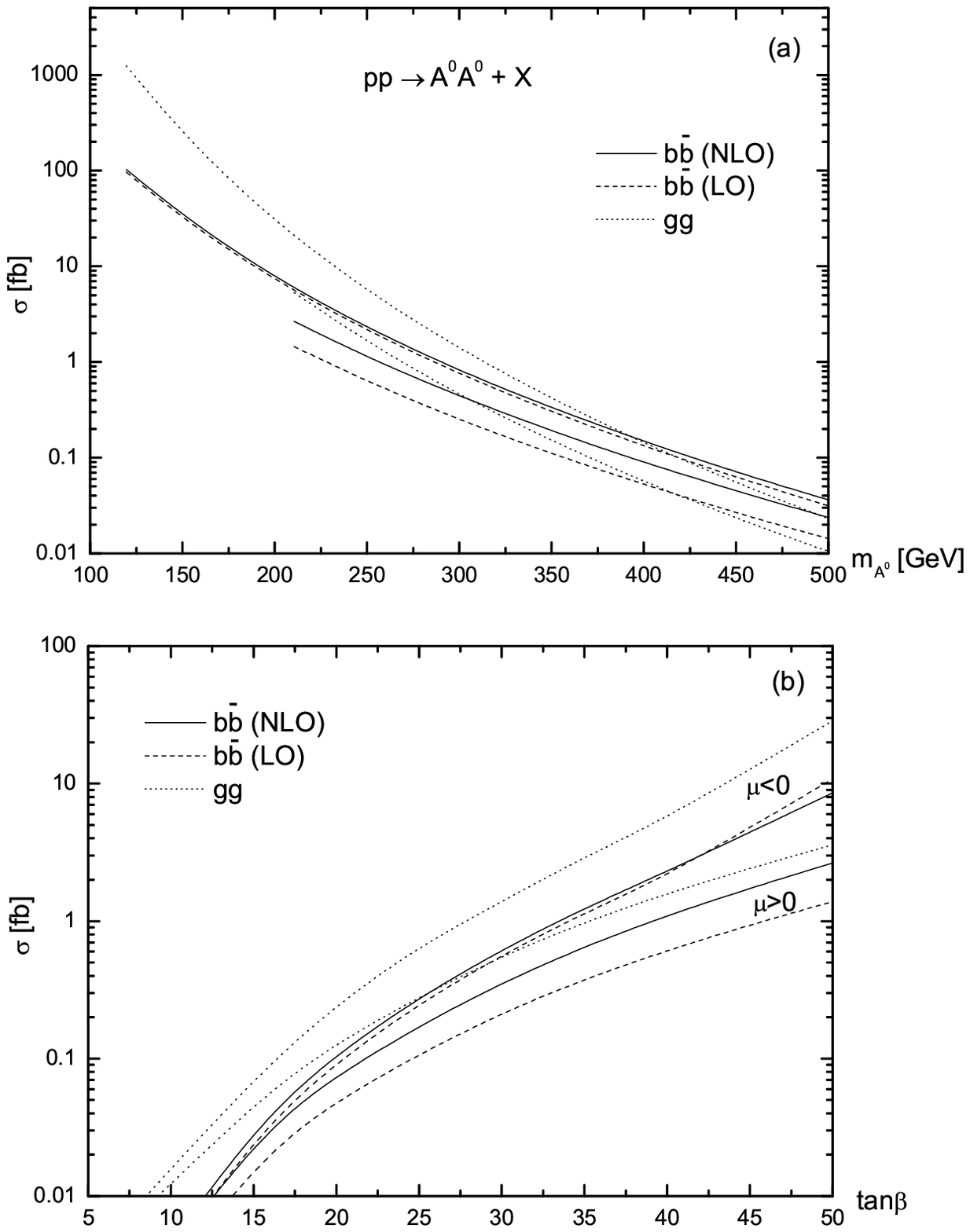,width=350pt}} \caption[]{Total
cross sections for the $A^0A^0$ production at the LHC (a) as
functions of $m_{A^0}$ for $\tan\beta=40$, $\mu<0$ (starting from
$m_{A^0}=120$ GeV) and $\tan\beta=40$, $\mu>0$ (starting from
$m_{A^0}=210$ GeV), and (b) as functions of $\tan\beta$ for
$m_{A^0}=250$ GeV, assuming $m_{1/2}=170$ GeV and $A_0=200$ GeV.
\label{fig:aa}}
\end{figure}

\begin{figure}[h!]
\centerline{\epsfig{file=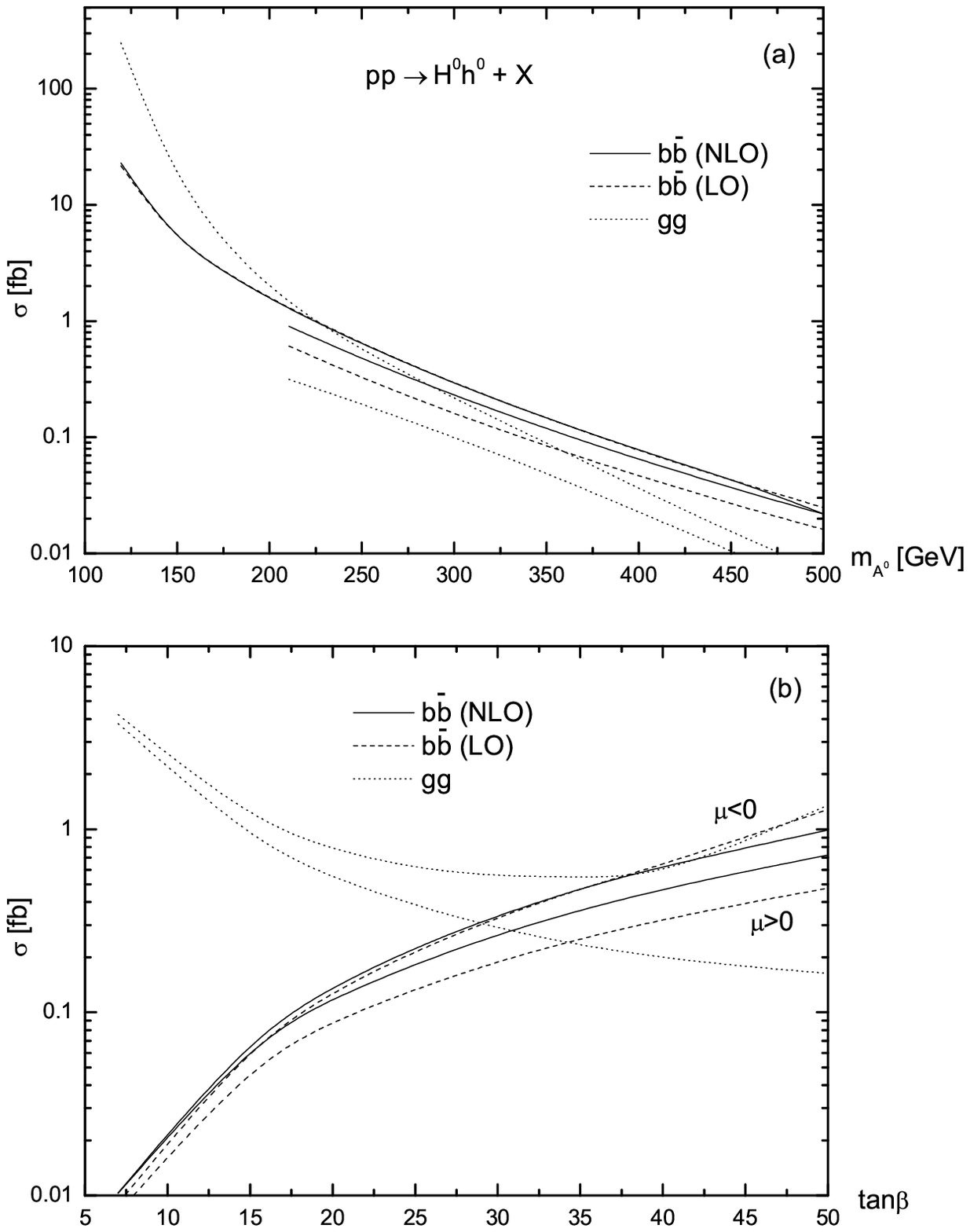,width=350pt}} \caption[]{Total
cross sections for the $H^0h^0$ production at the LHC (a) as
functions of $m_{A^0}$ for $\tan\beta=40$, $\mu<0$ (starting from
$m_{A^0}=120$ GeV) and $\tan\beta=40$, $\mu>0$ (starting from
$m_{A^0}=210$ GeV), and (b) as functions of $\tan\beta$ for
$m_{A^0}=250$ GeV, assuming $m_{1/2}=170$ GeV and $A_0=200$ GeV.
\label{fig:hhl}}
\end{figure}

In Figs.~\ref{fig:hlhl}--\ref{fig:hhl}, we give the total cross
sections for $h^0h^0$, $H^0H^0$, $A^0A^0$ and $h^0H^0$ production,
respectively, and compare the $b\bar b$-annihilation contributions
with the $gg$-fusion contributions \cite{hpair0,hpair1,
hpair11,hpair2}, which arise from quark and squark loops. In the
case of $h^0h^0$ production, Ref.~\cite{hpair2} indicated that
$b\bar b$ annihilation can be more important than $gg$ fusion for
large values of $\tan\beta$, but it is not so here (see
Fig.~\ref{fig:hlhl}(a)), which is due to the fact that we have
used a much larger decay width for the $H^0$ than the one in
Ref.~\cite{hpair2}. However, when $\tan\beta$ is small ($<15$),
the contributions of $b\bar b$ annihilation still can exceed those
of $gg$ fusion (see Fig.~\ref{fig:hlhl}(b)). As for $H^0H^0$
(Fig.~\ref{fig:hh}) and $A^0A^0$ (Fig.~\ref{fig:aa}) production,
$b\bar b$ annihilation is suppressed by a factor between 2 and 3
in most of the parameter space compared to $gg$ fusion except for
$m_{A^0}\geq 400$ GeV and $\tan \beta =40$, where the
contributions of $b\bar b$ annihilation are larger than those of
$gg$ fusion, but the corresponding cross sections are very small
($<0.1$ fb). In the case of $H^0h^0$ production, $b\bar b$
annihilation dominates for $m_{A^0}>250$ GeV and large values of
$\tan\beta$. From Figs.~\ref{fig:hlhl}--\ref{fig:hhl}, we also see
that the NLO QCD corrections to the total cross sections for these
four processes can enhance the LO results significantly for
$\mu>0$, generally by a few tens of percent, while for $\mu<0$,
the corrections are relatively small, and are even negative in
some parameter regions.

\begin{figure}[h!]
\centerline{\epsfig{file=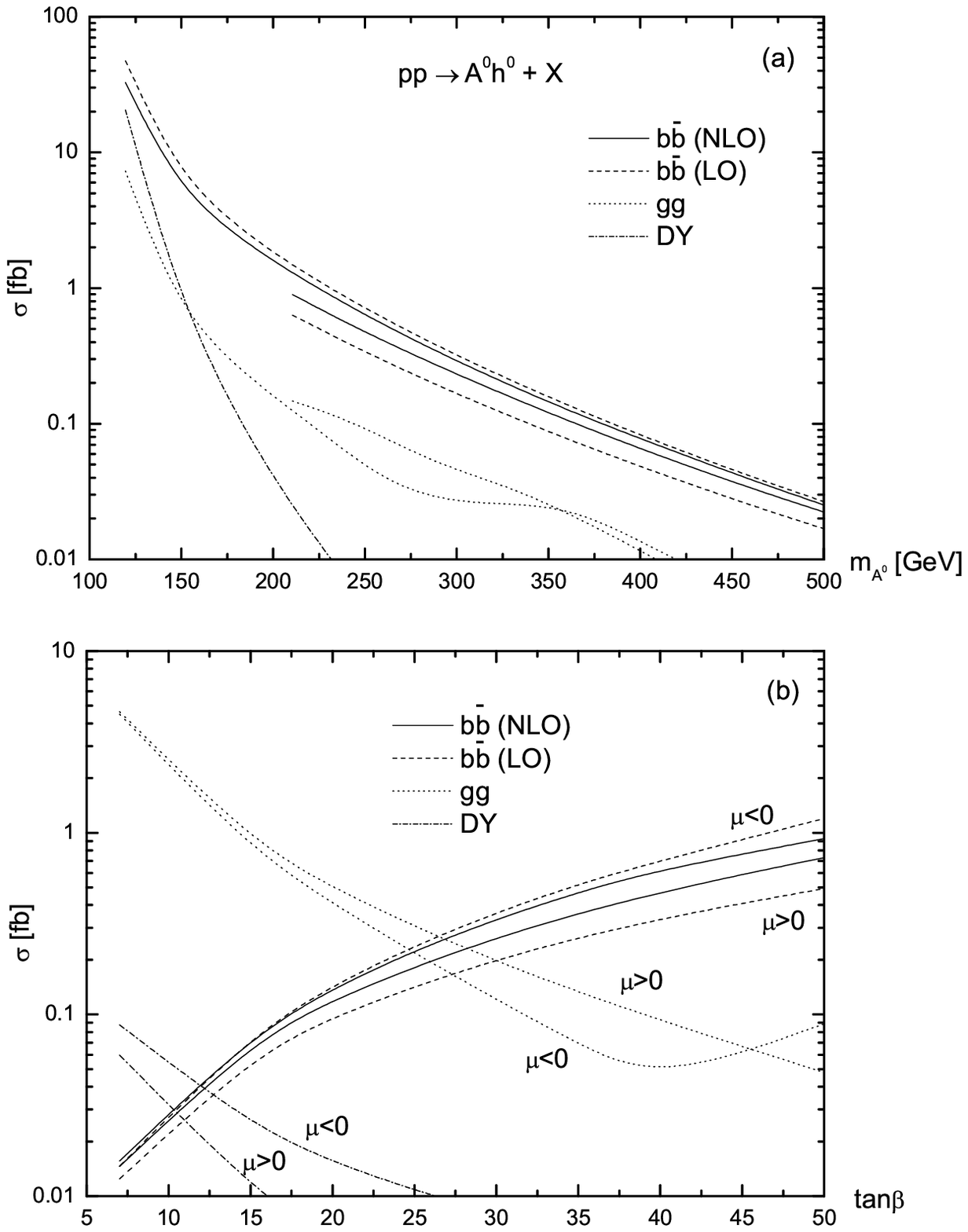,width=350pt}} \caption[]{Total
cross sections for the $A^0h^0$ production at the LHC (a) as
functions of $m_{A^0}$ for $\tan\beta=40$, $\mu<0$ (starting from
$m_{A^0}=120$ GeV) and $\tan\beta=40$, $\mu>0$ (starting from
$m_{A^0}=210$ GeV), and (b) as functions of $\tan\beta$ for
$m_{A^0}=250$ GeV, assuming $m_{1/2}=170$ GeV and $A_0=200$ GeV.
\label{fig:ahl}}
\end{figure}

\begin{figure}[h!]
\centerline{\epsfig{file=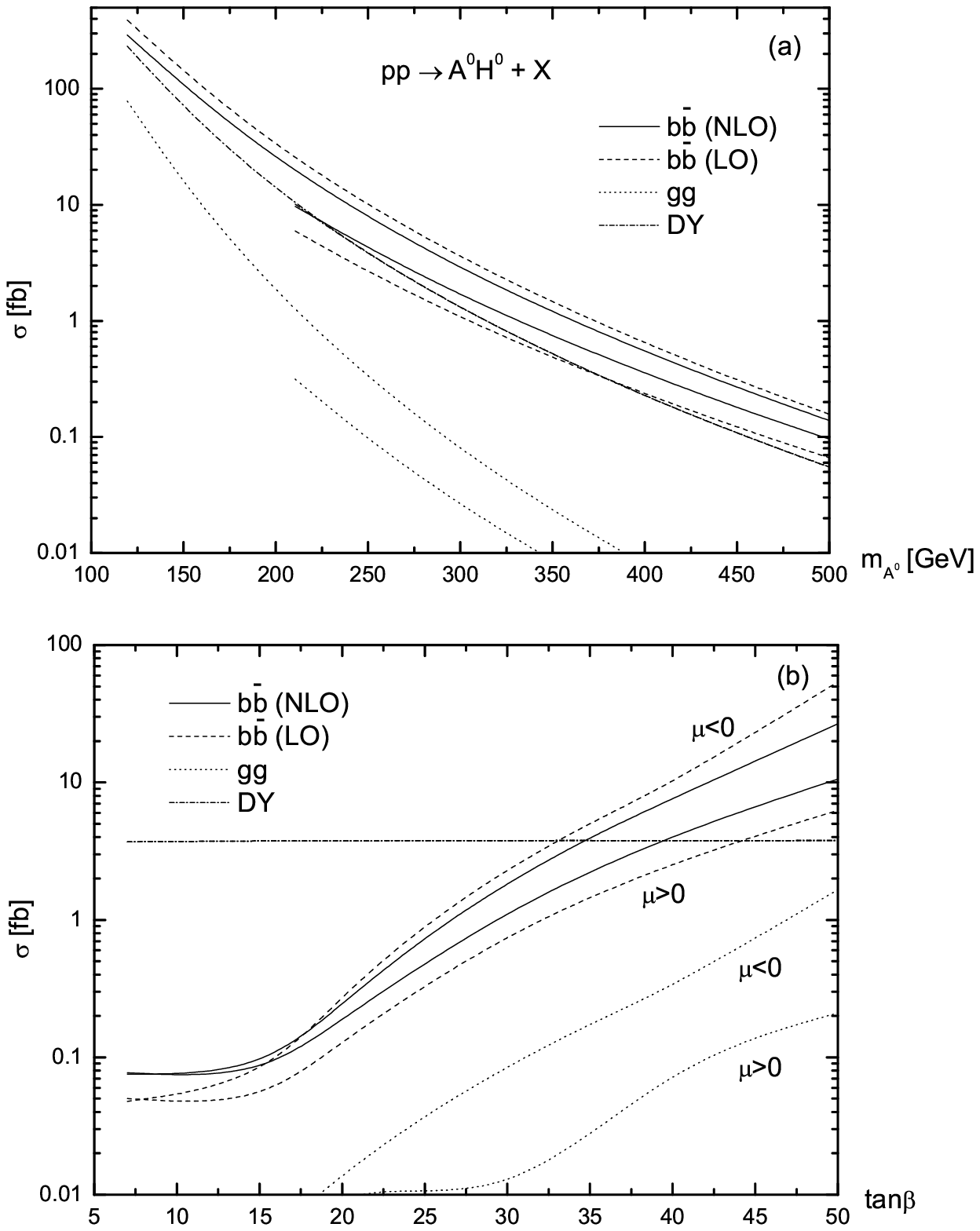,width=350pt}} \caption[]{Total
cross sections for the $A^0H^0$ production at the LHC (a) as
functions of $m_{A^0}$ for $\tan\beta=40$, $\mu<0$ (starting from
$m_{A^0}=120$ GeV) and $\tan\beta=40$, $\mu>0$ (starting from
$m_{A^0}=210$ GeV), and (b) as functions of $\tan\beta$ for
$m_{A^0}=250$ GeV, assuming $m_{1/2}=170$ GeV and $A_0=200$ GeV.
\label{fig:ah}}
\end{figure}

In Figs.~\ref{fig:ahl} and \ref{fig:ah}, we plot the total cross
sections for $pp\rightarrow A^0h^0, \ A^0H^0$ at the LHC as
functions of $m_{A^0}$ and $\tan\beta$, and compare the $b\bar b$
annihilation contributions with $gg$ fusion \cite{hpair0,hpair1,
hpair11,hpair2} and $q \bar q$ annihilation \cite{hpair0}. In the
case of $A^0h^0$ production, the $b \bar b$-annihilation
contributions dominate for large values of $\tan\beta$. For
example, when $\tan\beta=40$ and $m_{A^0}>250$ GeV (see
Fig.~\ref{fig:ahl}(a)), the contributions are several times larger
than $gg$ fusion contributions, and at least two orders of
magnitude larger than $q\bar q$ annihilation contributions.
However, for small values of $\tan\beta$, $gg$ fusion dominates,
as shown in Fig.~\ref{fig:ahl}(b). In the case of $A^0H^0$
production, after including the NLO QCD corrections, the cross
section for $b \bar b$ annihilation is lager than those of the
other two mechanisms for large values of $\tan\beta$ and most
values of $m_{A^0}$. Moreover, the $b\bar b$-annihilation
contributions can exceed 100 fb for $\tan\beta =40$ and
$m_{A^0}<150$ GeV, as shown in Fig.~\ref{fig:ah}(a). From
Figs.~\ref{fig:ahl} and \ref{fig:ah}, we also see that the NLO QCD
corrections to the total cross section for these two processes can
enhance the LO results significantly for $\mu>0$, generally by a
few tens of percent, while for $\mu<0$, the corrections are
negative and relatively small.

\begin{figure}[h!]
\centerline{\epsfig{file=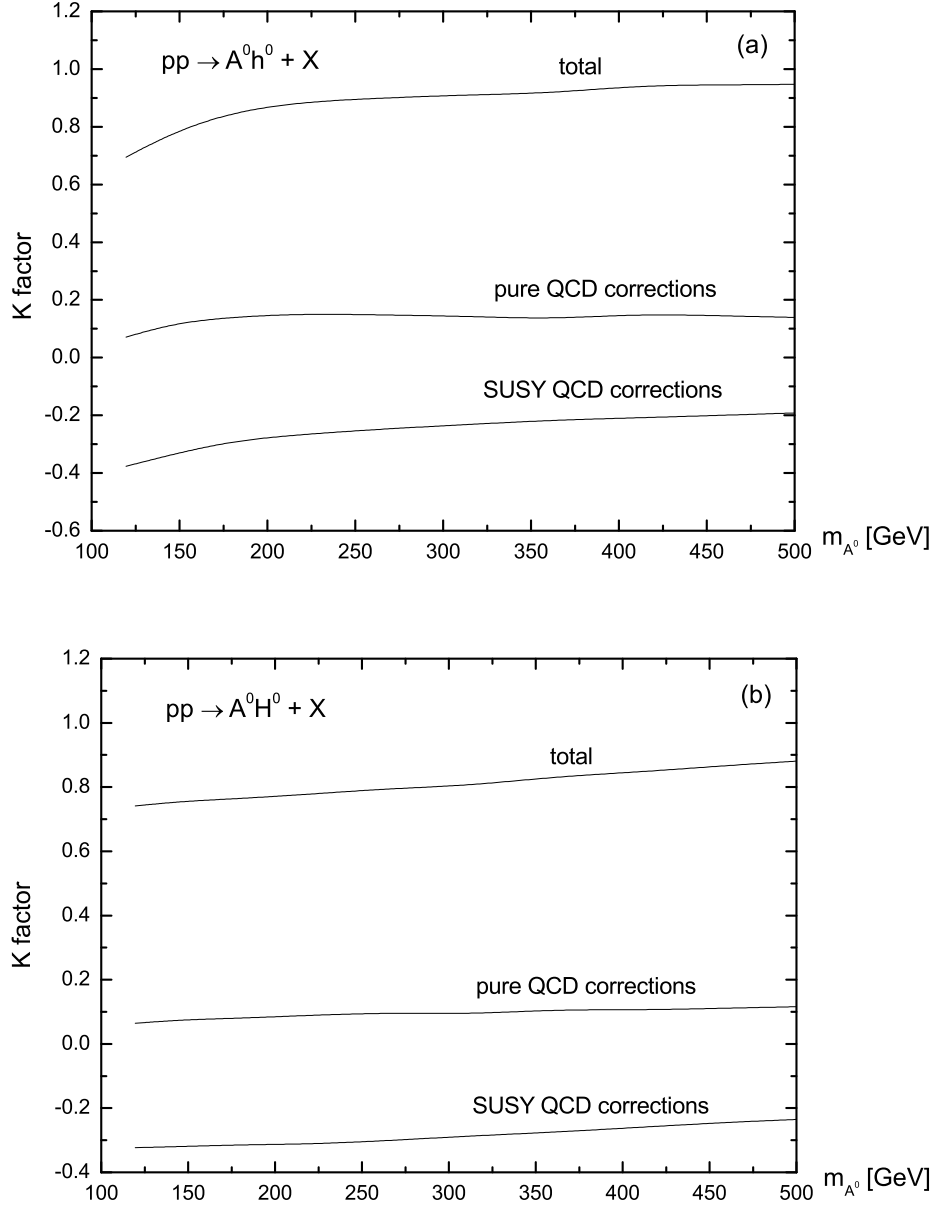,width=350pt}} \caption[]{$K=
\sigma_{NLO}/\sigma_{LO}$ for the $A^0h^0/A^0H^0$ production
through $b\bar b$ annihilation at the LHC as a function of
$m_{A^0}$, assuming $\tan\beta=40$, $m_{1/2}=170$ GeV, $A_0=200$
GeV and $\mu<0$. \label{fig:k}}
\end{figure}

Fig.~\ref{fig:k} gives the dependence of the ratio $K$ (defined as
the ratio of the NLO total cross sections to the LO ones) on
$m_{A^0}$ for $A^0h^0/A^0H^0$ production through $b\bar b$
annihilation based on the results shown in Fig.~\ref{fig:ahl}(a)
and \ref{fig:ah}(a). We see that in general the ratio $K$  is
negative, and becomes larger with the increasing $m_{A^0}$. For
example, when $m_{A^0}$ varies from 120 GeV to 500 GeV, the ratio
$K$ increases from 0.7 to 0.95 for $A^0h^0$ production
(Fig.~\ref{fig:k}(a)), and from 0.75 to 0.88 for $A^0H^0$
production (Fig.~\ref{fig:k}(b)). The contributions to the ratio
$K$ come from three IR finite parts: the LO total cross sections,
the pure QCD corrections and the SUSY-QCD virtual corrections, the
latter two of which are also shown in the figure.

\begin{figure}[h!]
\centerline{\epsfig{file=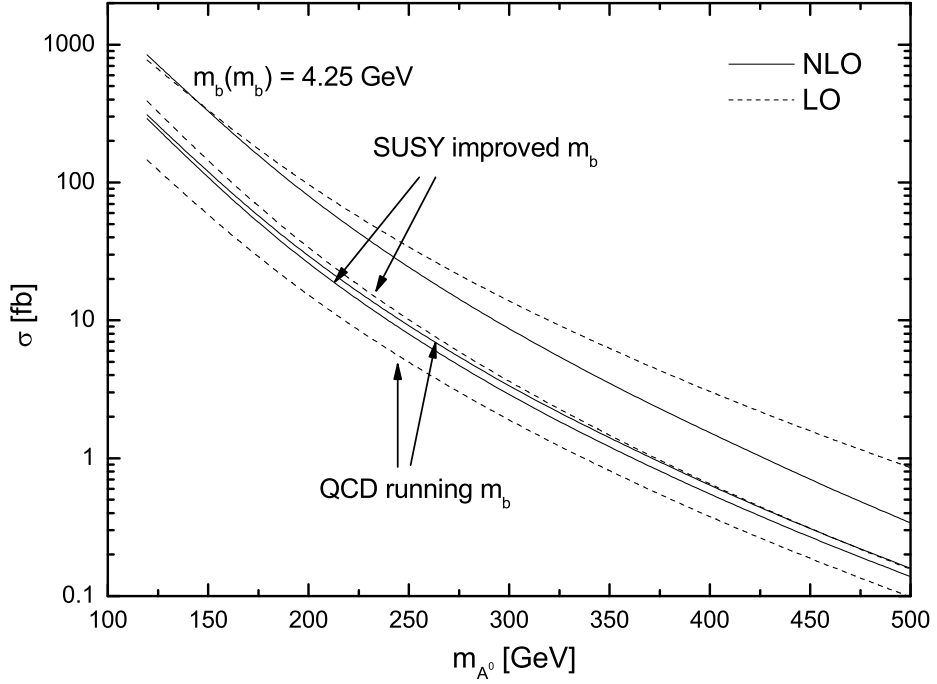,width=350pt}}
\caption[]{Dependence of the total cross sections for the $A^0H^0$
production through $b\bar b$ annihilation at the LHC on $m_{A^0}$
and $m_b$, assuming $\tan\beta=40$, $m_{1/2}=170$ GeV, $A_0=200$
GeV and $\mu<0$. \label{fig:mb}}
\end{figure}

The main parts of $b\bar b$ annihilation contributions for large
values of $\tan\beta$ originate from the Yukawa coupling
$b-b-\phi$, so the results are sensitive to the bottom quark mass.
In Fig.~\ref{fig:mb}, we show the effects of the choices of $m_b$
on the total cross section for $A^0H^0$ production through $b\bar
b$ annihilation, assuming $\tan\beta=40$, $m_{1/2}=170$ GeV,
$A_0=200$ GeV and $\mu<0$. We considered three different bottom
quark masses, (i) $\overline{\rm MS}$ bottom quark mass at the
scale of the $\overline{\rm MS}$ mass, i.e. $m_b(m_b)=4.25$ GeV,
(ii) QCD running bottom quark mass $m_b(\mu_r)$, and (iii) QCD
running plus SUSY improved bottom quark running mass. They can
have very different values. For example, in this figure, when
$m_{A^0}=250$ GeV, they are 4.25 GeV, 2.69 GeV and 3.17 GeV, which
leads to the LO total cross sections being 12.5 fb, 5 fb and 9.8
fb, respectively. Moreover, from Fig.~\ref{fig:mb}, we see that
the NLO QCD corrections can be very large in the cases of
$m_b(m_b)$ and QCD running $m_b$. For this reason we use the SUSY
improved bottom quark running mass, which improves the convergence
of the perturbation calculations, especially for large values of
$\tan\beta$, as shown in Fig.~\ref{fig:mb}.

\begin{figure}[h!]
\centerline{\epsfig{file=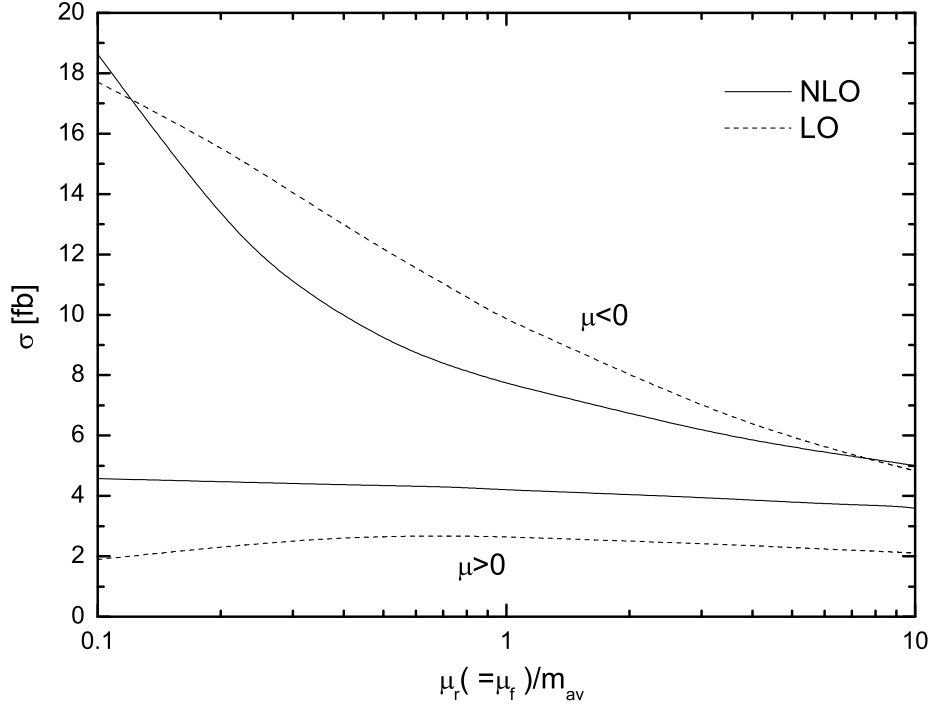,width=350pt}}
\caption[]{Dependence of the total cross sections for the $A^0H^0$
production through $b\bar b$ annihilation at the LHC on
renormalization/factorization scale, assuming $\tan\beta=40$,
$m_{1/2}=170$ GeV, $A_0=200$ GeV and $\mu<0$. \label{fig:mu}}
\end{figure}

Fig.~\ref{fig:mu} gives the dependence of the total cross section
for $A^0H^0$ production through $b\bar b$ annihilation at the LHC
on the renormalization/factorization scale for $\mu_r=\mu_f$. In
the case of $\mu>0$, the scale dependence of both the LO and the
NLO total cross sections is relatively weak. And for $\mu<0$, the
scale dependence of the total cross sections is reduced when going
from LO to NLO. For example, the cross sections vary by $\pm 20\%$
at LO but by $\pm 13\%$ at NLO in the region $0.5<\mu_f/m_{\rm
av}<2.0$.

\begin{figure}[h!]
\centerline{\epsfig{file=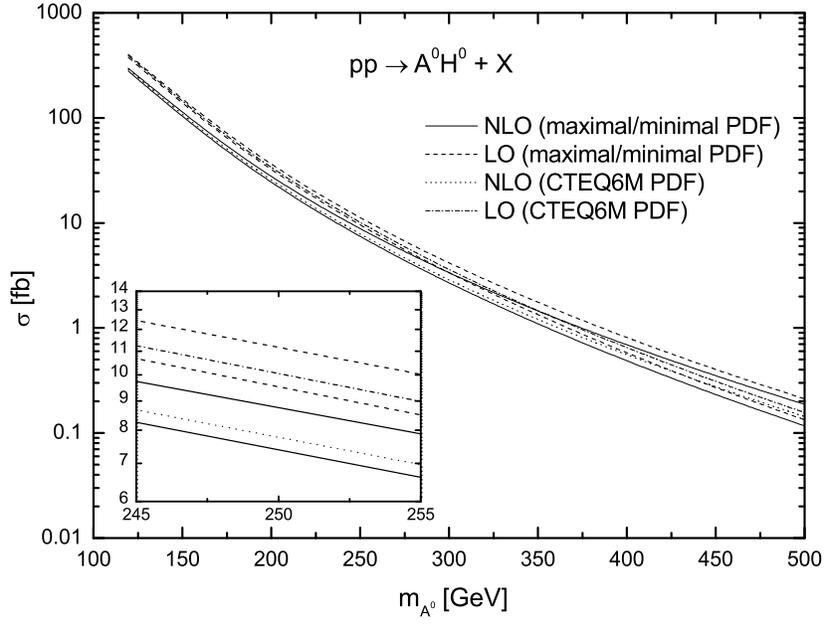,width=350pt}} \caption[]{The
PDF dependence of the total cross sections for the $A^0H^0$
production through $b\bar b$ annihilation at the LHC, assuming
$\tan\beta=40$, $m_{1/2}=170$ GeV, $A_0=200$ GeV and $\mu<0$.
\label{fig:uncert}}
\end{figure}

Since another source of uncertainty arises from the different
choice of PDFs, in Fig.~\ref{fig:uncert} we show the total cross
sections for $A^0H^0$ production through $b\bar b$ annihilation as
functions of $m_{A^0}$ for three different PDFs. We first use the
41 CTEQ6.1 PDF sets \cite{61cteq} to estimate the uncertainty in
the LO total cross sections. The LO results using the CTEQ6M PDFs
lie between the maximum and the minimum. The NLO total cross
sections are then calculated using three different PDF sets, one
of which is CTEQ6M, and the other two are the ones that gave the
maximum and minimum LO uncertainties. Observe that in this case
the uncertainty arising from the choice of PDFs increases with the
increasing $m_{A^0}$. Moreover, the dependence of the total cross
sections on PDFs is not decreased from LO to NLO.

\begin{figure}[h!]
\centerline{\epsfig{file=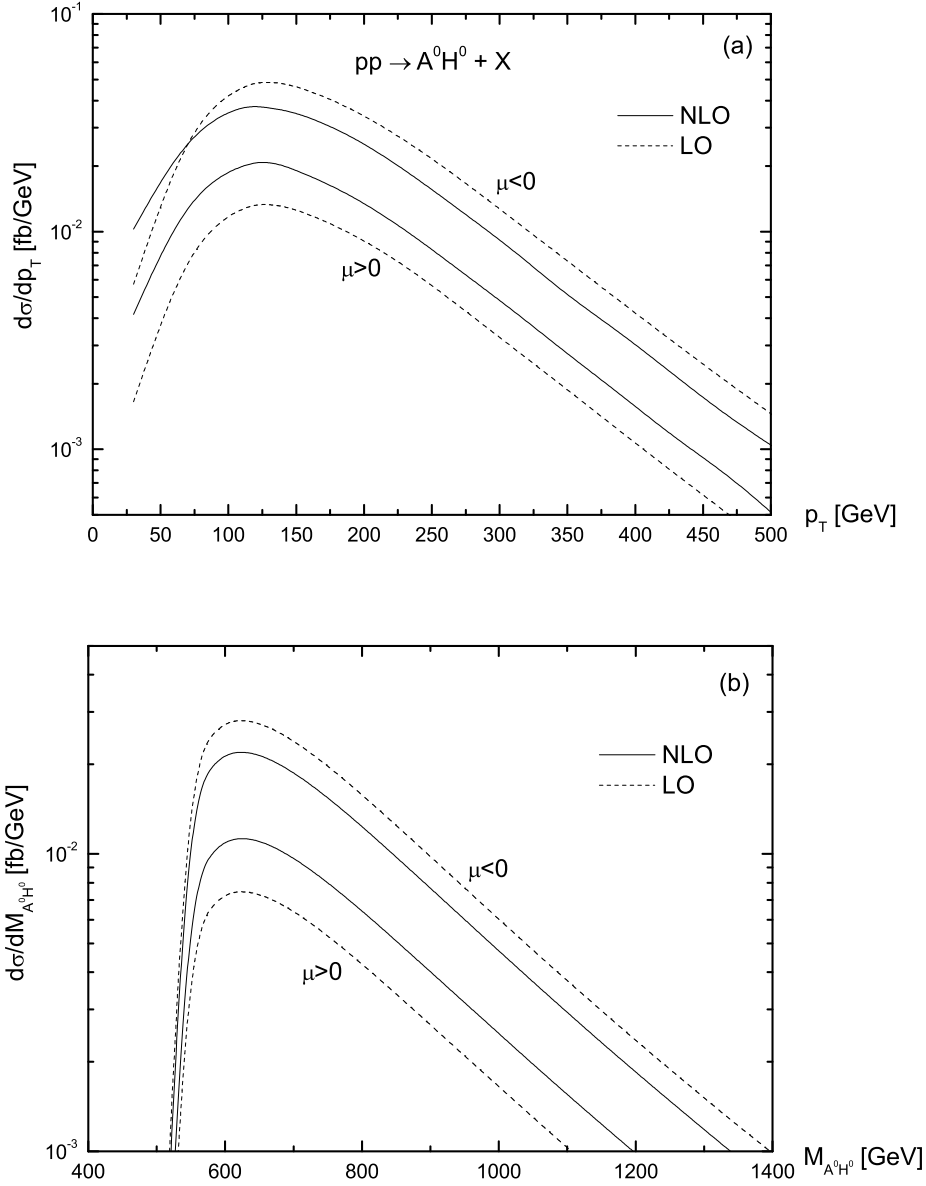,width=350pt}}
\caption[]{Differential cross sections in the transverse momentum
$p_T$ of $A^0$ and the invariant mass $M_{A^0H^0}$ for the
$A^0H^0$ production through $b\bar b$ annihilation at the LHC,
assuming $m_{A^0}=250$ GeV, $\tan\beta=40$, $m_{1/2}=170$ GeV and
$A_0=200$ GeV. \label{fig:pt}}
\end{figure}

In Fig.~\ref{fig:pt}, we display differential cross sections as
the functions of the transverse momentum $p_T$ of the $A^0$ and
the invariant mass $M_{A^0H^0}$, which are given by
Eqs.(\ref{integralpt}) and (\ref{invariant}), respectively, for
the $A^0H^0$ production through $b\bar b$ annihilation. In the
case of $\mu<0$, we find that the NLO QCD corrections reduce the
LO differential cross sections except for low $p_T$, while in the
case of $\mu>0$, the corrections always enhance the LO results.

In conclusion, we have calculated the complete NLO inclusive total
cross sections for pair production of neutral Higgs bosons through
$b\bar b$ annihilation in the MSSM at the LHC. In our
calculations, we used both the DREG scheme and the DRED scheme and
found that the NLO total cross sections in the above two schemes
are the same. Our results show that the $b\bar b$ annihilation
contributions can exceed  those of $gg$ fusion and $q\bar q$
annihilation for $h^0H^0$, $A^0h^0$ and $A^0H^0$ production when
$\tan\beta$ is large. For $\mu>0$ the NLO corrections enhance the
LO total cross sections significantly, and can reach a few tens
percent, while for $\mu<0$ the corrections are relatively small
and  negative in most of parameter space. Moreover, the NLO QCD
corrections reduce the dependence of these total cross sections on
the renormalization/factorization scale, especially for $\mu<0$.
We also used the CTEQ6.1 PDF sets to estimate the uncertainty in
both the LO and NLO total cross sections, and found that the
uncertainty arising from the choice of PDFs increases with the
increasing $m_{A^0}$.

\begin{acknowledgments}
We thank Cao Qin Hong, Han Tao and C.-P. Yuan for useful
discussions. This work is supported by China Postdoctoral Science
Foundation, the National Natural Science Foundation of China and
Specialized Research Fund for the Doctoral Program of Higher
Education and the U.S. Department of Energy, Division of High
Energy Physics, under Grant No.DE-FG02-91-ER4086.
\end{acknowledgments}

\appendix

\section{}
In this appendix, we give the relevant Feynman couplings
\cite{higgshunter}.

1. $H_i-b-\bar{b}:$ $igY_b(a_iP_L +a_i^\ast P_R)$
\begin{eqnarray*}
a_1=-\frac{1}{\sqrt{2}} \cos\alpha, \ \ \ \ \
a_2=\frac{1}{\sqrt{2}} \sin\alpha, \ \ \ \ \
a_3=-\frac{i}{\sqrt{2}} \sin\beta, \ \ \ \ \
a_4=\frac{i}{\sqrt{2}} \cos\beta,
\end{eqnarray*}
where $\alpha$ is the mixing angle in the CP-even neutral Higgs
boson sector.

2. $H_j-Z-A^0:$ $gZ^H_j(p_{A^0}-p_{H_j})^\mu/(2c_W)$
\begin{eqnarray*}
Z^H_1=-\sin(\beta-\alpha), \ \ \ \ \ Z^H_2=\cos(\beta-\alpha), \ \
\ \ \ Z^H_3=Z^H_4=0.
\end{eqnarray*}
Here we define the outgoing four--momenta of $A^0$ and $H_j$
positive.

3. $H_k-H_i-H_j:$ $igm_Z C_{kij}/(2c_W)$
\begin{eqnarray*}
&& C_{111}=-3\cos2\alpha \cos(\alpha+\beta), \qquad
C_{112}=2\sin2\alpha \cos(\alpha+\beta) + \sin(\alpha+\beta)
\cos2\alpha,
\\
&& C_{122}=-2\sin2\alpha \sin(\alpha+\beta) +\cos(\alpha+\beta)
\cos2\alpha, \qquad C_{133}=\cos2\beta \cos(\alpha+\beta),
\\
&& C_{134}=\sin2\beta \cos(\alpha+\beta), \qquad
C_{144}=-\cos2\beta \cos(\alpha+\beta),
\\
&& C_{222}=-3\cos2\alpha \sin(\alpha+\beta), \qquad
C_{233}=-\cos2\beta \sin(\alpha+\beta),
\\
&& C_{244}=\cos2\beta \sin(\alpha+\beta), \qquad
C_{234}=-\sin2\beta \sin(\alpha+\beta).
\end{eqnarray*}
The indexes $i,j$ and $k$ of $C_{kij}$ are symmetric, and other
coefficients are zero.

4. $H_k-\tilde{b}_l-\tilde{b}_m$: $igG^{k}_{lm}\equiv
ig[R^{\tilde{b}}\hat{G}^{k}(R^{\tilde{b}})^T]_{lm}$
\begin{eqnarray*}
\hat{G}^{1}= \left(\begin{array}{cc} -\frac{m_{Z}}{c_W}
\cos(\alpha+\beta) C_{bL} -\sqrt{2}m_bY_b
\cos\alpha & -\frac{1}{\sqrt{2}}Y_b (A_b \cos\alpha -\mu \sin\alpha) \\
-\frac{1}{\sqrt{2}} Y_b(A_b \cos\alpha -\mu \sin\alpha ) &
\frac{m_{Z}}{c_W} \cos(\alpha+\beta) C_{bR} -\sqrt{2}m_bY_b
\cos\alpha
\end{array} \right),
\end{eqnarray*}
\begin{eqnarray*}
\hat{G}^{2}= \left(\begin{array}{cc} \frac{m_{Z}}{c_W}
\sin(\alpha+\beta) C_{bL}+\sqrt{2}m_bY_b \sin\alpha &
\frac{1}{\sqrt{2}}Y_b(A_b \sin\alpha +\mu \cos\alpha)
\\
\frac{1}{\sqrt{2}}Y_b (A_b \sin\alpha +\mu \cos\alpha ) &
-\frac{m_{Z}}{c_W}\sin(\alpha+\beta)C_{bR}
+\sqrt{2}m_bY_b\sin\alpha
 \end{array} \right),
\end{eqnarray*}
\begin{eqnarray*}
\hat{G}^{3}=i\frac{m_b}{2m_W} \left(\begin{array}{cc} 0
& -A_b \tan\beta  -\mu \\
A_b \tan\beta +\mu & 0 \end{array} \right),
\end{eqnarray*}
\begin{eqnarray*}
\hat{G}^{4}=i\frac{m_b}{2m_W} \left(\begin{array}{cc} 0
& A_b -\mu\tan\beta \\
-A_b +\mu \tan\beta & 0 \end{array} \right).
\end{eqnarray*}

5. $Z^0-\tilde{b}_l-\tilde{b}_m$: $-igZ^{\tilde b}_{ml}
(p_{\tilde{b}_m}+p_{\tilde{b}_l})^\mu/c_W$
\begin{eqnarray*}
Z^{\tilde b}_{ml}= R^{\tilde b}_{m1} R^{\tilde b}_{l1} C_{bL} +
R^{\tilde b}_{m2} R^{\tilde b}_{l2} C_{bR},
\end{eqnarray*}
where $p_{\tilde{b}_m}$ and $p_{\tilde{b}_l}$ are the
four--momenta of $\tilde{b}_m$ and $\tilde{b}_m$ in direction of
the charge flow.

6. $H_i-H_j-\tilde{b}_l-\tilde{b}_m$: $ig^2G^{ij}_{lm}\equiv
ig^2[R^{\tilde{b}}\hat{G}^{ij}(R^{\tilde{b}})^T]_{lm}$
\begin{eqnarray*}
&& \hat{G}^{11}_{lm}= - \frac{1}{2}\left
(\frac{\cos2\alpha}{c_W^2} C_{bL} +\frac{m_b^2\cos^2
\alpha}{m_W^2\cos^2\beta} \right) \delta_{l1}\delta_{m1}
+\frac{1}{2}\left (\frac{\cos2\alpha}{c_W^2} C_{bR}
-\frac{m_b^2\cos^2\alpha}{m_W^2\cos^2\beta} \right)
\delta_{l2}\delta_{m2},
\\
&& \hat{G}^{22}_{lm}=\frac{1}{2}\left (\frac{\cos2\alpha}{c_W^2}
C_{bL} -\frac{m_b^2\sin^2\alpha}{m_W^2\cos^2\beta} \right)
\delta_{l1}\delta_{m1} -\frac{1}{2}\left
(\frac{\cos2\alpha}{c_W^2} C_{bR}
+\frac{m_b^2\sin^2\alpha}{m_W^2\cos^2\beta} \right)
\delta_{l2}\delta_{m2},
\\
&& \hat{G}^{33}_{lm}=\frac{1}{2} \left (\frac{\cos2\beta}{c_W^2}
C_{bL} -\frac{m_b^2\tan^2\beta}{m_W^2} \right)
\delta_{l1}\delta_{m1} -\frac{1}{2} \left(\frac{\cos2\beta}{c_W^2}
C_{bR} +\frac{m_b^2\tan^2\beta}{m_W^2} \right)
\delta_{l2}\delta_{m2},
\\
&& \hat{G}^{44}_{lm}= -\frac{1}{2}\left (\frac{\cos2\beta}{c_W^2}
C_{bL} +\frac{m_b^2}{m_W^2} \right ) \delta_{l1}\delta_{m1}
+\frac{1}{2}\left (\frac{\cos2\beta}{c_W^2} C_{bR}
-\frac{m_b^2}{m_W^2}\right ) \delta_{l2}\delta_{m2},
\\
&& \hat{G}^{12}_{lm}=\frac{\sin2\alpha}{2} \left [
\left(\frac{C_{bL}}{c_W^2} +\frac{m_b^2}{2m_W^2\cos^2\beta}\right)
\delta_{l1}\delta_{m1} - \left (\frac{C_{bR}}{c_W^2}
-\frac{m_b^2}{2m_W^2\cos^2\beta}\right)
\delta_{l2}\delta_{m2}\right],
\\
&& \hat{G}^{34}_{lm}=\frac{\sin2\beta}{2}\left [
\left(\frac{C_{bL}}{c_W^2} +\frac{m_b^2}{2m_W^2\cos^2\beta}\right
) \delta_{l1}\delta_{m1} -\left (\frac{C_{bR}}{c_W^2}
-\frac{m_b^2}{2m_W^2\cos^2\beta}\right)
\delta_{l2}\delta_{m2}\right].
\end{eqnarray*}
The indexes $i$ and $j$ of $G^{ij}_{lm}$ are symmetric, and other
coefficients are zero.

\section{}
In this appendix, we collect the explicit expressions for the
nonzero form factors in Eq.(\ref{formf}). For simplicity, we
introduce the following abbreviations for the Passarino-Veltman
integrals, which are defined as in Ref.~\cite{denner} except that
we take internal masses squared as arguments:
\begin{eqnarray*}
&& B_0^{a(\alpha)}=B_0(\alpha,0,0),
\\
&& B_0^{b(\alpha)}=B_0(\alpha, m_{\tilde g}^2, m_{\tilde b_l}^2),
\\
&& C_{p(q)}^{a}=C_{p(q)}(0,0,s,0,0,0),
\\
&& C_{p(q)}^{b(\alpha)}=C_{p(q)}(0,\alpha, 0 , m_{\tilde g}^2,
m_{\tilde b_l}^2, m_{\tilde b_m}^2),
\\
&& C_{p(q)}^{c(\alpha r)} =C_{p(q)}( 0,m_{H_r}^2, \alpha, 0, 0,
0),
\\
&& C_{p(q)}^{d(\alpha r)} =C_{p(q)}(0, m_{H_r}^2, \alpha,
m_{\tilde g}^2, m_{\tilde b_m}^2, m_{\tilde b_l}^2),
\\
&& C_{p(q)}^{e(\alpha)}=C_{p(q)}(\alpha, m_{H_i}^2, m_{H_j}^2, 0,
0, 0),
\\
&& D_{p(q)}^{a(\alpha)} =D_{p(q)}(\alpha, m_{H_j}^2, s, 0, 0,
m_{H_i}^2, 0, 0, 0, 0),
\\
&& D_{p(q)}^{b(\alpha)}= D_{p(q)}(0, s, m_{H_i}^2, \alpha, 0,
m_{H_j}^2, m_{\tilde g}^2, m_{\tilde b_l}^2, m_{\tilde b_m}^2,
m_{\tilde b_n}^2).
\end{eqnarray*}
Many functions above contain soft and collinear singularities, but
all the Passarino-Veltman integrals can be reduced to the scalar
functions $B_0$, $C_0$ and $D_0$. Here we present the explicit
expressions for $C_0$ and $D_0$, which contain the singularities,
and were used in our calculations:
\begin{eqnarray*}
&& C_{0}^{a}=\frac{C_\epsilon}{s} \bigg[\frac{1}{\epsilon^2}
-\frac{2 \pi^2}{3}\bigg],
\\
&& C^{c(\alpha r)}_0  =\frac{C_\epsilon}{\alpha -m_{H_r}^2}
\bigg[\frac{1}{\epsilon}\ln\bigg(\frac{-\alpha}{m_{H_r}^2}\bigg)
+\frac{1}{2}\ln^2\bigg(\frac{s}{m_{H_r}^2}\bigg)
-\frac{1}{2}\ln^2\bigg(\frac{s}{-\alpha} \bigg)
-\frac{\pi^2}{2}\bigg],
\\
&& D^{a(\alpha)}_0= \frac{C_\epsilon}{s\alpha}
\bigg[\frac{1}{\epsilon^2} +\frac{2}{\epsilon}
\ln\bigg(\frac{m_{H_i}m_{H_j}}{-\alpha}\bigg)
+\frac{2\pi^2}{3}\bigg]  -\frac{2C_\epsilon}{s\alpha}\Bigg \{{\rm
Li}_2\bigg(\frac{s +\alpha -m^2_{H_i}}{s}\bigg) -{\rm
Li}_2\bigg(\frac{s- m_{H_i}^2}{s}\bigg)
\\
&& \hspace{1.1cm} -{\rm
Li}_2\bigg[\frac{-s\alpha}{(s-m_{H_i}^2)(m_{H_i}^2-\alpha)}\bigg]
+{\rm Li}_2\bigg(\frac{-\alpha}{s-m_{H_i}^2}\bigg)+{\rm
Li}_2\bigg(\frac{m^2_{H_j}}{m^2_{H_j}-\alpha}\bigg)
\\
&& \hspace{1.1cm}
-\frac{1}{2}\ln^2\bigg[\frac{-s\alpha}{(s-m_{H_i}^2)(m_{H_i}^2-\alpha)}\bigg]
+\frac{1}{2}\ln^2\bigg(\frac{-\alpha}{s-m_{H_i}^2}\bigg)
+\ln\bigg(\frac{m_{H_i}^2-\alpha}{s}\bigg)\ln\bigg(\frac{s +\alpha
-m^2_{H_i}}{-\alpha}\bigg)
\\
&& \hspace{1.1cm}
-\frac{1}{2}\ln\bigg(\frac{m_{H_i}^2-\alpha}{s}\bigg)\ln\bigg
(\frac{sm^2_{H_j}}{\alpha^2}\bigg)
-\frac{1}{4}\ln^2\bigg(\frac{sm^2_{H_j}}{\alpha^2}\bigg)
+\frac{1}{4}\ln^2\bigg(\frac{m^2_{H_i}}{s}\bigg)
\\
&& \hspace{1.1cm} +\frac{1}{2}\ln\bigg(\frac{m^2_{H_j}}{s}\bigg)
\ln\bigg(\frac{m_{H_i}^2-\alpha}{m^2_{H_i}}\bigg)
+\frac{1}{2}\ln^2\bigg(\frac{m^2_{H_j}}{m^2_{H_j}-\alpha}\bigg)
\Bigg \},
\end{eqnarray*}
where we define $C_\epsilon=(4\pi\mu^2_r/s)^
\epsilon\Gamma(1-\epsilon)/\Gamma(1-2\epsilon)$, and note the
explicit expression for $D^{a(\alpha)}_0$ is in agreement with the
one in Ref.~\cite{d0}.

There are the following relations between the form factors:
\begin{eqnarray*}
f^\alpha_{2,4}=f^\alpha_{1,3} (a_{i,j,k} \leftrightarrow
a_{i,j,k}^\ast, C_{bL}\leftrightarrow C_{bR}, R_{l1}^{\tilde b}
\leftrightarrow R_{l2}^{\tilde b}, R_{m1}^{\tilde b}
\leftrightarrow R_{m2}^{\tilde b}).
\end{eqnarray*}
Thus we will only present the explicit expressions of $f_1^\alpha$
and $f_3^\alpha$. Corresponding to diagrams (a)--(m) in
Fig.~\ref{fig:feyvirt}, the form factors are
\begin{eqnarray*}
&& f^a_1= \sum_{k=1}^4 \frac{m_Z C_{kij}}{c_W s_{H_k}} \{ Y_b
a_k[s(C_0^{a} +2C_{1}^{a} +(2-\epsilon)C_{12}^{a})
+4(\epsilon-2)C_{00}^{a} +1]
\\
&& \hspace{1.0cm} - \sum_{l,m=1}^2 m_{\tilde g} G_{lm}^k
R_{m1}^{\tilde b} R_{l2}^{\tilde b} C_0^{b(s)}\},
\\
&& f^b_1 = \sum_{l,m=1}^2 \frac{i m_{\tilde g}Z_j^H Z_{lm}^{\tilde
b} R_{m1}^{\tilde b} R_{l2}^{\tilde b}}{c_W^2 (s - m_Z^2)} [(u -
t)(C_0^{b(s)} + C_1^{b(s)} +C_2^{b(s)}) + (m_{H_j}^2 -
m_{H_i}^2)(C_1^{b(s)} - C_2^{b(s)}) ],
\\
&& f^b_3 = \frac{- i Z_j^H}{c_W^2 (s - m_Z^2)}\{ C_{bL}
[s(2C_0^{a} +4C_{1}^{a} +2(1-\epsilon) C_{12}^{a}) -4(1-\epsilon)
C_{00}^{a} -1]
\\
&& \hspace{1.0cm} + \sum_{l,m=1}^2 4 R_{l1}^{\tilde b}
R_{m1}^{\tilde{b}} Z_{lm}^{\tilde b}C_{00}^{b(s)}\},
\\
&& f^c_1 +f^e_1 = -\sum_{l,m=1}^22Y_b (a_j G_{lm}^i R^{\tilde
b}_{l1} R^{\tilde b}_{m1}C_2^{d(ti)} +a_i G_{ml}^j R^{\tilde
b}_{l2} R^{\tilde b}_{m2}C_2^{d(tj)}),
\\
&& f^c_3 +f^e_3= \frac{2Y_b}{t}\{Y_b a_ia_j^\ast \sum_{r=i,j}[
t(C_0^{c(tr)} + C_1^{c(tr)} +2C_{12}^{c(tr)} +2C_{22}^{c(tr)}
+3C_2^{c(tr)}) - t\epsilon (C_{12}^{c(tr)}
\\
&& \hspace{1.5cm}  +C_{22}^{c(tr)} +C_{2}^{c(tr)}) +4(2-\epsilon)
C_{00}^{c(tr)} -m_{H_r}^2 (C_{0}^{c(tr)} +C_{1}^{c(tr)} +(2
-\epsilon)C_{12}^{c(tr)}+C_{2}^{c(tr)})
\\
&& \hspace{1.5cm}   -1] + \sum_{l,m=1}^2 m_{\tilde g}(a_j^\ast
G^i_{lm} R^{\tilde b}_{l2} R^{\tilde b}_{m1}C_0^{d(ti)} + a_i
G^j_{ml}R^{\tilde b}_{m1} R^{\tilde b}_{l2}C_0^{d(tj)})\},
\\
&& f_1^d + f_1^f = - \sum_{l,m=1}^2 2Y_b ( a_j G^i_{ml} R^{\tilde
b}_{l2} R^{\tilde b}_{m2}C_2^{d(ui)} + a_i G^j_{lm} R^{\tilde
b}_{l1} R^{\tilde b}_{m1} C_2^{d(uj)}),
\\
&& f_3^d + f_3^f= -\frac{2 Y_b}{u}  \{Y_b a_j a_i^\ast
\sum_{r=i,j}[ u(C_0^{c(ur)} + C_1^{c(ur)} +2C_{12}^{c(ur)}
+2C_{22}^{c(ur)} +3C_2^{c(ur)}) - u\epsilon (C_{12}^{c(ur)}
\\
&& \hspace{1.5cm}  +C_{22}^{c(ur)} +C_{2}^{c(ur)}) +4(2-\epsilon)
C_{00}^{c(ur)} -m_{H_r}^2 (C_{0}^{c(ur)} +C_{1}^{c(ur)} +(2
-\epsilon)C_{12}^{c(ur)}+C_{2}^{c(ur)})
\\
&& \hspace{1.5cm}  -1] + \sum_{l,m=1}^2m_{\tilde g}[a_i^\ast
G^j_{lm}  R^{\tilde b}_{l2} R^{\tilde b}_{m1}C_0^{d(uj)} + a_j
 G^i_{ml} R^{\tilde b}_{l1} R^{\tilde b}_{m2}C_0^{d(ui)}]\},
\\
&& f_1^g +f_1^h = - \sum_{l=1}^2 2 Y_b^2 a_i a_j R^{\tilde b}_{l1}
R^{\tilde b}_{l2} \frac{ m_{\tilde g}}{tu}(t B_0^{b(u)}+ u
B_0^{b(t)}),
\\
&& f_3^g +f_3^h = Y_b^2 \{ \frac{1 - \epsilon}{tu} ( t a_j
a_i^\ast B_0^{a(u)} -  u a_i a_j^\ast B_0^{a(t)}) + \sum_{l=1}^2
(R^{\tilde b}_{l2})^2 [ \frac{a_i a_j^\ast }{t^2} ((m_{\tilde
g}^2- m_{\tilde{b}_l}^2)(B_0^{b(0)}
\\
&& \hspace{1.5cm} - B_0^{b(t)}) - t B_0^{b(t)} ) + \frac{a_j
a_i^\ast}{u^2}((m_{\tilde g}^2 - m_{\tilde{b}_l}^2)(B_0^{b(u)}-
B_0^{b(0)}) + u B_0^{b(u)})]\},
\\
&& f^i_3 = 2Y_b^2a_i a_j^\ast[C_0^{c(ti)} + C_0^{c(tj)} -
(1+\epsilon)C_0^{e(s)} - sD_0^{a(t)} + (u - t\epsilon)D_1^{a(t)}],
\\
&& f^j_1 = - \sum_{l,m,n=1}^2 2  m_{\tilde g} G^i_{nm} G^j_{ln}
R^{\tilde b}_{l2} R^{\tilde b}_{m1} D_0^{b(t)},
\\
&& f^j_3 = \sum_{l,m,n=1}^2 2 G^i_{nm} G^j_{ln}R^{\tilde b}_{l1}
R^{\tilde b}_{m1} D_3^{b(t)},
\\
&& f^k_3 = - 2 Y_b^2 a_j a_i^\ast [C_0^{c(ui)} + C_0^{c(uj)}
-(1+\epsilon) C_0^{e(s)} - s D_0^{a(u)} + (t
-u\epsilon)D_1^{a(u)}],
\\
&& f^l_1 = - \sum_{l,m,n=1}^2 2  m_{\tilde g} G^i_{mn} G^j_{nl}
R^{\tilde b}_{m2} R^{\tilde b}_{l1} D_0^{b(u)},
\\
&& f^l_3 = - \sum_{l,m,n=1}^2 2 G^i_{mn} G^j_{nl} R^{\tilde
b}_{l1}R^{\tilde b}_{m1} D_3^{b(u)} ,
\\
&& f^m_1 = \sum_{l,m=1}^2 2  m_{\tilde g} G^{ij}_{lm} R^{\tilde
b}_{l2} R^{\tilde b}_{m1} C_0^{b(s)}.
\end{eqnarray*}

\section{}
In this Appendix, we collect the explicit expressions for the
amplitudes squared for the radiation of a real gluon.The results
for massless $b(\bar b)$ emission can be obtained by crossing
symmetry. Since these expressions are only used for the hard
non-collinear parts of the real corrections in Eq.(\ref{nonHC})
and Eq.(\ref{qHC}), they have no singularities and can be
calculated in $n=4$ dimensions.

For simplicity, we define the following invariants:
\begin{eqnarray*}
&& s_3=(p_2 +p_3)^2, \qquad s_4=(p_1+p_3)^2, \qquad s_5=(p_1
+p_2)^2,
\\
&& t=(p_b -p_2)^2, \qquad t'=(p_b-p_3)^2, \qquad u=(p_a -p_2)^2,
\\
&& u'=(p_a-p_3)^2, \qquad u_6=(p_b -p_1)^2, \qquad
u_7=(p_a-p_1)^2,
\end{eqnarray*}.
Then for real gluon emission we find
\begin{eqnarray*}
\overline{|M^{b\bar b}_{ij}|}^2 = g^4 g_s^2
\left(\overline{|M^{(HH)}_{ij}|}^2 + \overline{|M^{(ZZ)}_{ij}|}^2
+ \overline{|M^{(Zt)}_{ij}|}^2 + \overline{|M^{(Zu)}_{ij}|}^2 +
\overline{|M^{(tt)}_{ij}|}^2 +\overline{|M^{(tu)}_{ij}|}^2
+\overline{|M^{(uu)}_{ij}|}^2 \right),
\end{eqnarray*}
with
\begin{eqnarray*}
&& \overline{|M^{(HH)}_{ij}|}^2 = \frac{2 m_Z^2 Y_b^2}{c_W^2}
\sum_{k,l=1}^4 (a_k a_l^\ast + a_k^\ast a_l) C_{kij} C_{lij}
\frac{(s + t' - u')(s - t' + u')}{t' u' s_{H_k} s_{H_l}^\ast},
\\
&& \overline{|M^{(ZZ)}_{ij}|}^2 = \frac{\delta_{i3} (Z_j^H)^2
(C_{bL}^2 + C_{bR}^2)}{2c_W^4 (s_5 - m_Z^2)^2 t'u'} \{s_4 u_6(s +
t') + s_4 u_7(s +u') + u_6 u_7(2 s  +t' +u')- t' u_7^2
\\
&& \hspace{1.0cm} - u_6^2u' - m_{H_i}^2[ t'(s_4 + t' + 2u_6 - u_7)
+ u'(u' + s_4 - u_6 + 2u_7) + s(2s + 2s_4 + 2t'
\\
&& \hspace{1.0cm} + 3u_6 + 3u_7 + 2u') + m_{H_i}^2(4s + t' +
u')]\},
\\
&& \overline{|M^{(Zt)}_{ij}|}^2 = \frac{i\delta_{i3} Y_b^2
Z_j^H}{c_W^2(s_5-m_Z^2)} (C_{bR}a_j a_i^\ast + C_{bL}a_i a_j^\ast)
\{ \frac{2}{tt'u_7}(s + t' + u_6 - m_{H_i}^2)[m_{H_i}^2(2s + 2s_4
+ t'
\\
&& \hspace{1.0cm} + 2(u_6 + 2u_7 + u') - 4m_{H_i}^2)-(2s_4 + t' +
2u_6)u_7]
\\
&& \hspace{1.0cm} - \frac{1}{t'u_7}[ss_4 + u_7(4s_4 + t') - u_6u'
- m_{H_i}^2(s + 4s_4 + t' + 4u_7 + 3u' -4m_{H_i}^2)]
\\
&& \hspace{1.0cm} - \frac{2}{tu_7u'}(u_7 - m_{H_i}^2)[(2s_4 + 2u_7
+ 3u' - 4m_{H_i}^2) (m_{H_i}^2 - u_6)-(s + t')(u' - 2m_{H_i}^2)]
\\
&& \hspace{1.0cm} + \frac{1}{t'u_7u'}[2u_6 - 2u_7 - u')(u_6u'
-t'u_7 + (t' - u')m_{H_i}^2) + s(3u' + m_{H_i}^2(4s + 4t'
\\
&& \hspace{1.0cm} + 6(u_6 + u_7) -4u_6u_7  -8m_{H_i}^2) + s_4(u' -
2u_6 - 2u_7 + 4m_{H_i}^2))]
\\
&& \hspace{1.0cm} - \frac{1}{tu'}[s_3(s + 4t) - t'u + tu' -
m_{H_j}^2(s + 4s_3 + 4t + 3t'  + u' - 4m_{H_j}^2)]
\\
&& \hspace{1.0cm} + \frac{1}{tt'u'}[s(s_3(t' -2t - 2u) - 4tu) +
(2t + t'  - 2u)(tu' - t'u ) + m_{H_j}^2((2t + t'
\\
&& \hspace{1.0cm} - 2u)(t' - u') + s(4s + 4s_3 + 6t + 3t' + 6u +
4u' - 8m_{H_j}^2))]\},
\\
&& \overline{|M^{(tt)}_{ij}|}^2 = 16Y_b^4 |a_i|^2 |a_j|^2
\{\frac{2}{tt'u_7^2}(s + t' + u_6 - m_{H_i}^2)[u_7(s_4 + u_6) -
m_{H_i}^2(s + s_4 + u_6 + 2u_7
\\
&& \hspace{1.0cm} + u' - 2m_{H_i}^2)] - \frac{1}{tt'u_7u'}[(u_6 -
u_7 - u')(u_6u' - t'u_7 + m_{H_i}^2(t' - u'))
\\
&& \hspace{1.0cm} - s(2u_6u_7 + s_4(u_6 + u_7 - u')) +
sm_{H_i}^2(2s + 2s_4 + 2t' + 3u_6 + 3u_7 + u'- 4m_{H_i}^2)]
\\
&& \hspace{1.0cm} + \frac{2}{t^2u_7u'}(u_7 - m_{H_i}^2)[(s + t' +
2u_6 -2 m_{H_i}^2)(m_{H_i}^2 - u') + (m_{H_i}^2 -u_6)(s_4 + u_7)]
\\
&& \hspace{1.0cm} + \frac{1}{t^2u_7^2}[u_7((s + t')(s_4 + u_7) -
u_6u') - m_{H_i}^2(t'(s_4 + u' - m_{H_i}^2) + u_7(s + 2t' - u'))]
\\
&& \hspace{1.0cm} + \frac{1}{t^2u'}[s_3t - m_{H_j}^2(s_3 + t + t'
- m_{H_j}^2)] + \frac{1}{t'u_7^2}[s_4u_7 - m_{H_i}^2(s_4 + u_7 +
u' - m_{H_i}^2)]\},
\\
&& \overline{|M^{(tu)}_{ij}|}^2 = 16 Y_b^4 a_i^2 a_j^{\ast2} \{
\frac{1}{t u_6u'} [t'u_7 - u_6u'- (s_4 - m_{H_i}^2)(s + 2u_6) +
m_{H_i}^2(2s_4 + t' + u' - 2m_{H_i}^2)]
\\
&& \hspace{1.0cm} + \frac{2}{tuu_6u_7}m_{H_i}^2m_{H_j}^2 +
\frac{2}{tu_6u'u u_7}(tu_6 - m_{H_i}^2m_{H_j}^2)[u(u_7 -
m_{H_i}^2) + u_7(u - m_{H_i}^2)]
\\
&& \hspace{1.0cm} - \frac{1}{t'uu_7}[(s_4 - m_{H_i}^2)(s + 2u_7) +
t'u_7 - u_6u' - m_{H_i}^2(2s_4 + t' + u' - 2m_{H_i}^2)]
\\
&& \hspace{1.0cm} - \frac{2}{t'u u_6u_7}(u_6 -
m_{H_i}^2)[(m_{H_i}^2 - u_7)(s_4 + t' + u_6) + m_{H_i}^2(s + 2u_7
+ u' - 2m_{H_i}^2)]
\\
&& \hspace{1.0cm} + \frac{1}{t'u_6u_7u'} [(u_6 - u_7)(u_6u' -t'u_7
) -s(2u_6u_7 + s_4(u_6 + u_7)) + m_{H_i}^2((u_6 - u_7)(t' - u')
\\
&& \hspace{1.0cm} + s(2s + 2s_4 + 2t' + 3u_6 + 3u_7 + 2u'-
4m_{H_i}^2))] + \frac{2}{tt'uu_7}(t - m_{H_j}^2)(uu_7 -
m_{H_i}^2m_{H_j}^2)
\\
&& \hspace{1.0cm} + \frac{1}{tt'uu'} [(t - u)(t u' - t'u) -s(2tu +
s_3(t + u)) + m_{H_j}^2((t -  u)(t' - u') + s(2s
\\
&& \hspace{1.0cm} + 2s_3 + 3t + 2t' + 3u + 2u' - 4m_{H_j}^2))] \}.
\end{eqnarray*}

\newpage

\end{document}